\documentclass{aa}

\usepackage{amsmath}
\usepackage{amssymb}
\usepackage{natbib}
\usepackage{wasysym}
\usepackage{graphicx}
\usepackage{float}
\usepackage{epsfig}
\usepackage{booktabs}
\usepackage{multirow}
\usepackage{lineno}
\usepackage{slashbox}
\usepackage{etoolbox}
\usepackage[varg]{txfonts}

\bibpunct{(}{)}{;}{a}{}{,}

\begin{document}

\title{EUV-driven ionospheres and electron transport on extrasolar giant planets orbiting active stars}
\titlerunning{Ionospheres of Extrasolar Giant Planets orbiting active stars}

\author{J.M.~Chadney \inst{1}
	\and M.~Galand \inst{1}
	\and T.T.~Koskinen \inst{2}
	\and S.~Miller \inst{3}
	\and J.~Sanz-Forcada \inst{4}
	\and Y.C.~Unruh \inst{1}
	\and R.V.~Yelle \inst{2}
}

\institute{Department of Physics, Imperial College London, Prince Consort Road, London SW7 2AZ, UK
	\and Lunar and Planetary Laboratory, University of Arizona, 1629 E. University Blvd., Tucson, AZ 85721, USA
	\and Department of Physics and Astronomy, University College London, London WC1E 6BT, UK
	\and Centro de Astrobiolog\'{i}a (CSIC-INTA), ESAC Campus, P.O. Box 78, E-28691 Villanueva de la Ca\~{n}ada, Madrid, Spain
}

\abstract{The composition and structure of the upper atmospheres of extrasolar giant planets (EGPs) are affected by the high-energy spectrum of their host stars from soft X-rays to the extreme ultraviolet (EUV). This emission depends on the activity level of the star, which is primarily determined by its age. In this study, we focus upon EGPs orbiting K- and M-dwarf stars of different ages -- \object{$\epsilon$ Eridani}, \object{AD Leonis}, \object{AU Microscopii} -- and the \object{Sun}. XUV (combination of X-ray and EUV) spectra for these stars are constructed using a coronal model. These spectra are used to drive both a thermospheric model and an ionospheric model, providing densities of neutral and ion species. Ionisation -- as a result of stellar radiation deposition -- is included through photo-ionisation and electron-impact processes. The former is calculated by solving the Lambert-Beer law, while the latter is calculated from a supra-thermal electron transport model. We find that EGP ionospheres at all orbital distances considered (0.1-1 AU) and around all stars selected are dominated by the long-lived H$^+$ ion. In addition, planets with upper atmospheres where H$_2$ is not substantially dissociated (at large orbital distances) have a layer in which H$_3^+$ is the major ion at the base of the ionosphere. For fast-rotating planets, densities of short-lived H$_3^+$ undergo significant diurnal variations, with the maximum value being driven by the stellar X-ray flux. In contrast, densities of longer-lived H$^+$ show very little day/night variability and the magnitude is driven by the level of stellar EUV flux. The H$_3^+$ peak in EGPs with upper atmospheres where H$_2$ is dissociated (orbiting close to their star) under strong stellar illumination is pushed to altitudes below the homopause, where this ion is likely to be destroyed through reactions with heavy species (e.g. hydrocarbons, water). The inclusion of secondary ionisation processes produces significantly enhanced ion and electron densities at altitudes below the main EUV ionisation peak, as compared to models that do not include electron-impact ionisation. We estimate infrared emissions from H$_3^+$, and while, in an H/H$_2$/He atmosphere, these are larger from planets orbiting close to more active stars, they still appear too low to be detected with current observatories.}


\maketitle

\section{Introduction}
The ionospheres of solar system planets have been observed and modelled in great depth \citep[e.g.][]{Nagy2002,Witasse2008}. Visiting spacecraft making in situ measurements or remote observations of emission in the infrared (IR) and ultraviolet (UV), in combination with detailed modelling, have enabled us to glean at least a basic understanding of this layer of atmosphere throughout the solar system. This apparatus has not yet been fully applied to exoplanetary ionospheres. Many recent advances in technology and technique have enabled the detection of ions, such as C$^+$ and Si$^{2+}$, which have been detected during transits of the hot Jupiter HD209458b \citep{Vidal-Madjar2004,Linsky2010}. However, observations remain difficult and must be supplemented by models. A potential diagnostic tool would be the measurement of emissions in the IR from the H$_3^+$ ion. This has been performed for Jupiter, Saturn, and Uranus \citep[e.g.][]{Drossart1989,Baron1991,Trafton1993,Rego2000,Stallard2008a,Miller2010}, not only allowing determination of H$_3^+$ densities and temperatures, but also providing valuable constraints for ionospheric models \citep[e.g.][]{Moore2015}. However, H$_3^+$ emissions have never been detected from an exoplanetary atmosphere, and the predicted emission levels are too low to be detected from planets orbiting stars of similar type and age to the Sun \citep{Shkolnik2006,Koskinen2007a}.

The ionosphere can be important in regulating the stability of upper atmospheres on extrasolar giant planets (EGPs). For example, previous modelling suggests that IR-active species, such as the H$_3^+$ ion, can act as a thermostat in EGP atmospheres and prevent them from undergoing hydrodynamic escape. Hence there are two different regimes of atmospheric escape depending on orbital distance, stellar heating and dissociation of molecular coolants upon which the composition and structure of the upper atmosphere depends. At large orbital distances ($a > 0.2$~AU for a Jupiter-like planet orbiting a Sun-like star), the main thermal escape mechanism is Jeans escape and the atmosphere is in a stable state of hydrostatic equilibrium, whereas planets with small orbital distances ($a < 0.2$~AU around a Sun-like star) undergo hydrodynamic escape \citep{Koskinen2007a,Koskinen2014}. In addition, electrodynamics in the ionosphere can modulate escape rates and influence the structure of the upper atmosphere through ion drag and resistive heating \citep{Koskinen2014a}.

Ionisation in upper planetary atmospheres occurs through two main processes. Photo-ionisation (or primary ionisation) by stellar photons has been calculated in past EGP models \citep[e.g.][]{Yelle2004,GarciaMunoz2007,Koskinen2010,Koskinen2013a}. However, to our knowledge, a full description of secondary ionisation, by photo-electrons and their secondaries, has not been included in EGP models. In solar system gas giants, secondary ionisation has been shown to strongly affect the lower ionosphere as well as the main ionisation peak \citep[e.g.][]{Kim1994,Galand2009}.

In addition, previous EGP studies have used solar XUV (X-ray and extreme ultraviolet) fluxes as substitutes for stellar fluxes. Inter-stellar extinction makes it very difficult (or even impossible at certain wavelengths) to measure spectra of other stars in the extreme ultraviolet (EUV). However, in a recent study, \citet{Sanz-Forcada2011} calculated mass-loss rates for gas giants orbiting close to a number of different stars. This was achieved by determining synthetic XUV spectra of these stars using emission measure distributions (EMD).

In contrast to previous models that assumed a Sun-like host star, we focus on the ionospheres of EGPs around young and active K- and M-dwarf stars. A significant number of the known exoplanets orbit such stars, and the upper atmospheres of these planets can be very different from similar planets orbiting Sun-like stars \citep{Chadney2015}. For the first time, we use actual stellar XUV spectra, determined using a stellar coronal model for each star. In addition to photo-ionisation, we also consider electron-impact ionisation by using a rigorous model of the energy degradation of suprathermal electrons. This is particularly important for planets orbiting active stars that typically emit more X-rays and short-wavelength EUV radiation than the Sun. This radiation is capable of enhancing the ionisation rates by producing high-energy photo-electrons that extend the ionosphere to lower altitudes. Based on our calculations, we also provide predictions of the H$_3^+$ emission power for planets orbiting active stars to determine whether they are better targets for future observations than planets around Sun-like stars.

The work described in this paper builds on that from \citet{Chadney2015}, in which the neutral thermosphere and atmospheric escape in EGPs around stars of different type and activity level were studied. In that work a simple method of scaling the stellar EUV flux was developed, based upon the observed X-ray flux. It was also shown that the transition from slow Jeans escape to hydrodynamic escape occurs at significantly larger orbital distances around more active stars. In the current paper, we combine the stellar coronal model and EGP thermosphere model \citep{Chadney2015} with a newly constructed 1D ionosphere model, in particular to study the effect of secondary ionisation by photo-electrons in EGP ionospheres around active stars.

It is important to note that we focus here upon ionisation in the thermosphere (at pressures $p<10^{-6}$~bar), where stellar EUV and soft X-ray radiation is absorbed. While in solar system planets, the main ionospheric peak is located in the thermosphere, \citet{Koskinen2014a} showed that in close-orbiting EGPs, the photo-ionisation of metals, such as Na and K, creates a stronger peak at lower altitudes, in the stratosphere. The ratio of the stellar X-ray to EUV flux is higher in more active stars \citep{Chadney2015} and this can further enhance the electron densities below the EUV ionisation layer.

This paper is laid out as follows. Section~\ref{sec:models} provides descriptions of the various components of the model. The results of the model runs are presented in Sect.~\ref{sec:results}, including a study of the sensitivity of our results to the resolution of the H$_2$ photo-absorption cross section (Sect.~\ref{sec:sig_res}). Section~\ref{sec:prodrates} discusses primary and secondary ionisation rates. Density predictions are provided in Sects.~\ref{sec:stable} and \ref{sec:hydrodyn}, focussing on planets undergoing Jeans escape, and hydrodynamic escape respectively. In Sect.~\ref{sec:vary_a}, we discuss the influence of orbital distance on ionospheric densities. Section~\ref{sec:scaled_spec} presents the effects of using the scaled solar XUV spectra from \citet{Chadney2015} to represent stellar spectra from other low-mass stars. Finally, Sect.~\ref{sec:H3Pemissions} provides estimates of IR emissions from the H$_3^+$ ion for planets orbiting a variety of different stars, at various orbital distances.

\section{Models} \label{sec:models}
\begin{figure}
\centering
\includegraphics[width=.3\textwidth]{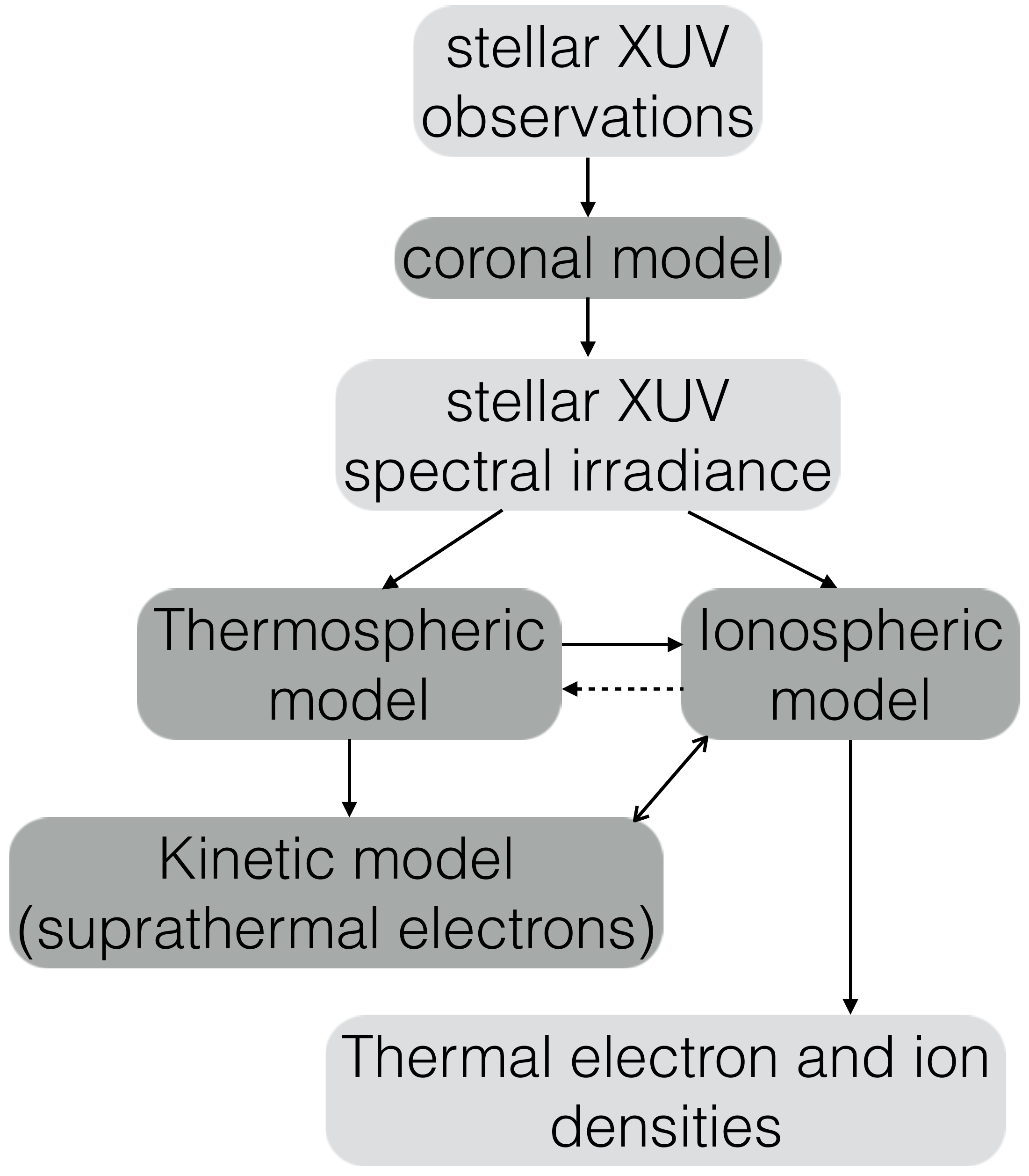}
\caption{Schematic of model elements (dark grey boxes) and inputs/outputs (light grey boxes).} 
\label{fig:model}
\end{figure}

We use a set of coupled models to describe the upper atmosphere of EGPs. The relationship between the different model elements is shown in the schematic in Fig.~\ref{fig:model}. The only source of incident energy considered is stellar XUV radiation. We incorporate accurate descriptions of both the high-energy stellar spectrum, by using a coronal model (see Sect.~\ref{sec:coronalmodel}) and an upper planetary atmospheric model that consists of coupled thermospheric (see Sect.~\ref{sec:thermomodel}), ionospheric, and suprathermal electron transport models (see Sect.~\ref{sec:ionomodel}). In all cases, we assume a rotational period for the planet of 10~hours, similar to that of Jupiter and Saturn. This is justified because we concentrate on planets orbiting at distances larger than 0.2~AU from their host stars. Taking 0.2~AU as the inner most orbital distance in this study allows us to sample planets in both escape regimes, whilst also ensuring that the assumption of a pure H$_2$/H/He neutral upper atmosphere is more likely to be correct.

\subsection{Stellar coronal model} \label{sec:coronalmodel}
We include in this study the K-dwarf $\epsilon$ Eridani ($T_{\text{eff}}=4900$~K) and the M-dwarfs AD Leonis ($T_{\text{eff}}=3370$~K) and AU Microscopii ($T_{\text{eff}}=3720$~K). These are close stars for which a good number of observations are available from Chandra, XMM-Newton, ROSAT, EUVE, and FUSE, which is important since these observations are used to calibrate the coronal model. This choice of stars also makes up a coarse parameter grid to study stars of different age and spectral type, where the Sun is not a valid proxy. $\epsilon$ Eridani has commonly been used as an analogue of the hot Jupiter host star HD189733, the two stars being of similar type, metallicity, and age \citep[e.g.][]{Moses2011,Venot2012}. AD Leonis has been used in previous studies of habitable planets \citep[e.g.][]{Tarter2007}. AU Microscopii is a well-known flare star that is very young and active \citep{Cully1993}. As a result of the lower contrast between star and planet, observations of planetary atmospheres of EGPs around M stars have recently received particular attention. Selecting two M stars with similar bolometric luminosity but very different X-ray and EUV luminosity allows us to gauge the likely range of EGP ionospheres for cool host stars with different activity levels.

Our knowledge of the stellar EUV spectra needed as input to our ionospheric model is patchy: the problem is that the high-energy stellar radiation that is absorbed in upper planetary atmospheres is also absorbed by the inter-stellar medium (ISM). Photons of $\lambda < 91.2$~nm are capable of ionising an atmosphere composed of hydrogen and helium. For almost all stars but the Sun, the wavelength region between about 40 and 91.2~nm is unobservable. Therefore, to obtain XUV (combination of X-ray and EUV) stellar irradiances, either a coronal model of the star or an appropriate scaling of the solar spectrum must be used, such as that described in \citet{Chadney2015}. We obtain accurate XUV spectra using a stellar coronal model \citep{Sanz-Forcada2002,Sanz-Forcada2003,Sanz-Forcada2011}. These spectra are obtained by constructing an emission measure distribution (EMD) of the corona, transition region and chromosphere of each star. A more detailed description of the construction of the XUV spectra for each of the three selected stars is provided in \citet{Chadney2015}.
 
\subsection{Thermospheric model} \label{sec:thermomodel}
To calculate the number density, velocity, and temperture profiles in the upper atmosphere of a planet, we used the one-dimensional thermospheric EGP model developed by \citet{Koskinen2013a,Koskinen2013b,Koskinen2014}. In all simulations the planetary parameters of HD209458b were used (radius $R_p =$~1.32~$R_{\text{Jupiter}}$, mass $M_p =$~0.69~$M_{\text{Jupiter}}$). The model solves the vertical equations of motion from the 10$^{-6}$~bar level up to the exobase for a fluid composed of H, H$_2$, and He, and their associated ions H$^+$, H$_2^+$, H$_3^+$, He$^+$, and HeH$^+$. The model is driven by the stellar irradiance and determines profiles with altitude of neutral densities, bulk velocity, and neutral temperature. The profiles obtained are a global average, obtained by dividing the incoming stellar flux by a factor of 4.

It should be noted that the thermospheric model is not fully coupled to the ionospheric model. Considering ionospheric densities when calculating the neutral thermosphere is important, especially to properly treat IR cooling by the H$_3^+$ ion. Therefore the thermospheric model does include a calculation of ionospheric densities. However, this calculation is not as thorough as that included in the full ionospheric model (see Sect.~\ref{sec:ionomodel}), as electron-impact ionisation by supra-thermal electrons is not taken into account. Instead, heating of the thermosphere due to photo-electrons is included in the thermospheric model using a fixed heating efficiency of 93~\%. \citet{Koskinen2013a} showed that using a fixed heating efficiency is appropriate for altitudes below $3R_p$. The value of 93~\% that we used is valid for photo-electrons created by photons of energy up to 50 eV at an electron mixing ratio of 0.1 \citep{Cecchi-Pestellini2009}, which is attained near the temperature peak in our models. The principal aim of this study is to compare the differences in composition of the ionosphere of planets orbiting active stars with different high-energy spectral energy distributions, and not to precisely model the neutral temperature and density profiles in these atmospheres.

\subsection{Ionospheric model} \label{sec:ionomodel}
The ionospheric model constructed for this work solves the one-dimensional coupled continuity equations for the ions H$^+$, H$_2^+$, H$_3^+$, and He$^+$ to provide the densities of these ions. The new ionosphere model is required because the previous models did not include secondary ionisation by photo-electrons. In spherical coordinates, where only radial transport is considered, the continuity equation for each ion $i$ reduces to
\begin{equation} \label{eqn:cont_vert}
\frac{\partial n_i}{\partial t} + \frac{1}{r^2}\frac{\partial}{\partial r}\left(r^2 \Phi_i\right) = P_i - L_i,
\end{equation}
where $n_i$ is the number density of ion $i$, $P_i$ is the production rate, and $L_i$ is the loss rate of ion $i$, $\Phi_i=n_iv_i$ is the radial transport flux, $v_i$ being the drift velocity of ion $i$. Ion production and loss occur through the photo-chemical reactions listed in Table~\ref{tab:reactionsiono}. Neutral species are ionised through photo-ionisation and electron-impact ionisation. We included the latter by using a suprathermal electron transport model adapted to EGPs that is based on the solution to the Boltzmann equation with transport, angular scattering, and energy degradation of photo-electrons and their secondaries taken into account. For further information on the suprathermal electron transport model, see \citet{Moore2008a} and \citet{Galand2009}.

We assumed here that the source of incident energy is solar or stellar XUV radiation. No electron precipitation from the space environment was included. A number of different parameters are required as inputs to the ionospheric model; these are the solar or stellar spectrum, neutral temperature and densities, chemical reaction rates, and cross sections. We used the TIMED/SEE instrument for measurements of solar XUV irradiances and a coronal model (see Sect.~\ref{sec:coronalmodel}) to produce stellar XUV spectra. The neutral atmosphere was assumed to be a constant background and was obtained for a set of planetary parameters and stellar irradiances using a thermospheric model (see Sect.~\ref{sec:thermomodel}). The assumption that the neutral densities remain constant is valid as long as ion densities remain small. This is the case in the ionospheres of solar system bodies, and we have also verified that this assumption remains valid for all the cases considered in this study because of the relatively large orbital distances.

\begin{figure}
\centering
\includegraphics[width=.48\textwidth]{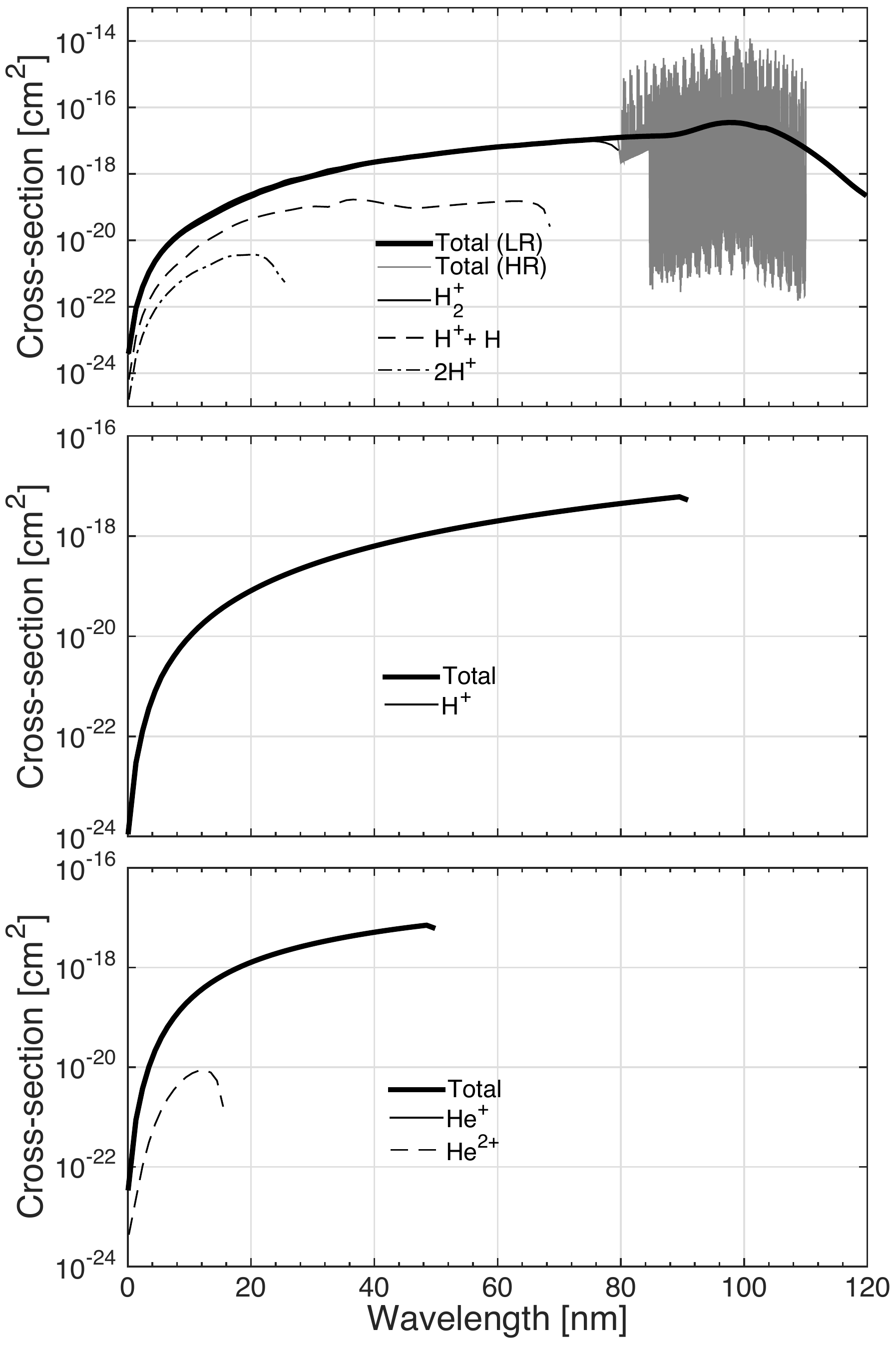}
\caption{Photo-absorption (thick black and thin grey lines) and photo-ionisation (thin black lines) cross sections used in the ionospheric model for the three neutral species considered and their associated ions. Panel (a) shows cross sections for H$_2$; in thick black we show the low-resolution (LR) photo-absorption cross section and in thin grey the high-resolution (HR) version (at 700~K) including structure beyond the ionisation threshold. The thin solid line plots the production of H$_2^+$, the dashed line the production of H$^+$+H, and the dashed-dotted line the production of 2H$^+$. Panel (b) shows the photo-absorption cross section of H as a thick black line, which is indistinguishable from the production of H$^+$ through photo-ionisation, which is shown as a thin solid line. Panel (c) shows the photo-absorption cross section of He as a thick black line, which is indistinguishable from the production of He$^+$, which is shown as a thin solid line, and the production He$^{2+}$ (the dashed line).}
\label{fig:cross-sec}
\end{figure}

Chemical reaction rates and cross sections are provided in Table~\ref{tab:reactionsiono} and Fig.~\ref{fig:cross-sec}, respectively. Cross sections of two different wavelength resolutions are used to describe the photo-absorption of H$_2$. At wavelengths greater than the ionisation threshold, H$_2$ absorbs in a large number of narrow lines, comprising the Lyman, Werner, and Rydberg bands. Typical ionosphere models use a low-resolution version of the H$_2$ cross section in these bands, including all of the previous exoplanet models. In Sect.~\ref{sec:sig_res} we present the effects on EGP atmospheres of using either low- or high-resolution versions of these cross sections.

The reaction of H$^+$ with vibrationally excited H$_2$ (reaction 11 in Table~\ref{tab:reactionsiono}) is an important loss process for H$^+$, but its reaction rate is not well constrained; as such, it is a large source of uncertainty in ionospheric models applied to giant planets. \citet{McElroy1973} was the first to note that this reaction would become exothermic for vibrationally excited H$_2$ and that there may well be enough excited H$_2$ in gas giant atmospheres to render this reaction significant. The rate coefficient of reaction 11 is known to be $1.0\times10^9$~cm$^3$s$^{-1}$ \citep{Huestis2008}. However, the proportion of vibrationally excited H$_2$ is unknown. We used the expression determined by \citet{Yelle2004} from Jupiter observations:
\begin{equation}
\frac{\left[\text{H}_2(\nu\geq4)\right]}{\left[\text{H}_2\right]} = \text{exp}(-2.19\times10^4/T).
\end{equation}
Therefore, in the ionospheric model, we assumed the following reaction:
\begin{equation*}
\text{H}^+ + \text{H}_2 \rightarrow \text{H}_2^+ + \text{H} \,\,\,\, \text{(reaction 11b)}
\end{equation*}
(note H$_2$ is in the base state), with the following reaction rate:
\begin{equation}
k_{11\text{b}} = 1.0\times10^9 \text{exp}(-2.19\times10^4/T)\,\text{cm}^3\text{s}^{-1}.
\end{equation}

Diurnal variations in rotating planets are determined by calculating ion densities with altitude at a given latitude and varying the stellar zenith angle over the course of a day. We neglected horizontal circulation and assumed that vertical gradients are dominant over horizontal gradients.

We assumed that any species heavier than He is confined below the lower boundary of the model, located at a pressure of 1~$\mu$bar. This pressure level was therefore assumed to correspond to the homopause. Species are diffusively separated above this level, heavier species being confined to lower altitudes. This assumption is valid if the eddy diffusion coefficient $K_{zz}$ is lower than $\sim10^2$ -- $10^3$~m$^2$~s$^{-1}$ at a pressure level of 1~$\mu$bar. The eddy diffusion coefficient is not well known in exoplanetary atmospheres, but its value is estimated to be about $10^2$~m$^2$~s$^{-1}$ at Jupiter \citep{Yelle1996}. $K_{zz}$ may well be higher in close-orbiting EGPs than at Jupiter \citep{Koskinen2010}, in which case the homopause could be located at a lower pressure than we assumed, leading to the destruction of the IR coolant H$_3^+$ through reactions with hydrocarbons, for instance. There are currently no constraints on $K_{zz}$ in EGPs however, a situation that might be improved by the detection of IR emissions from the H$_3^+$ ion (see Sect.~\ref{sec:H3Pemissions}).

Atmospheric escape could also cause heavier species to be present above the 1~$\mu$bar level. In certain planets undergoing hydrodynamic escape, hydrocarbons could be dragged upwards into the thermosphere by escaping hydrogen. We estimated in which of the cases discussed in this paper hydrocarbons could be present in the thermosphere because of atmospheric escape: it only occurs in EGPs orbiting within 0.2~AU from AD Leo (see Sect.~\ref{sec:vary_a}).

\begin{table*}
\centering
\small
\caption{Chemical reactions used in the ionospheric model. The reaction rates are in units of cm$^3$s$^{-1}$ for two-body reactions and cm$^6$s$^{-1}$ for three-body reactions. We use wavelength-dependent cross sections for photo-ionisation and electron-impact ionisation reactions.}
\begin{tabular}{llcl}
\toprule
\# & Reaction & Reaction rate & Reference \\
\midrule
 & Photo-ionisation: & & \\
1 & $\text{H}_2 + h\nu \rightarrow \text{H}_2^+ + e^-$ & \multicolumn{2}{l}{\citet{Backx1976,Kossmann1989},} \\
  &													   & \multicolumn{2}{l}{\citet{Chung1993,Yan1998}} \\
2 & $\text{H}_2 + h\nu \rightarrow \text{H}^+ + \text{H} + e^-$ & \multicolumn{2}{l}{\citet{Chung1993} and 2H$^+$ references} \\
3 & $\text{H}_2 + h\nu \rightarrow 2\text{H}^+ + 2e^-$ & \multicolumn{2}{l}{\citet{Dujardin1987,Kossmann1989a},} \\
  &													   & \multicolumn{2}{l}{\citet{Yan1998}} \\
4 & $\text{H} + h\nu \rightarrow \text{H}^+ + e^-$ & \multicolumn{2}{l}{\citet{Verner1996}} \\
5 & $\text{He} + h\nu \rightarrow \text{He}^+ + e^-$ & \multicolumn{2}{l}{\citet{Verner1996}} \\
 & & & \\
 & Electron-impact ionisation: & & \\
6 & $\text{H}_2 + e^- \rightarrow \text{H}_2^+ + e^- + e^-$ & \multicolumn{2}{|l}{\citet{VanWingerden1980,Ajello1991},} \\
7 & $\text{H}_2 + e^- \rightarrow \text{H}^+ + \text{H} + e^- + e^-$ & \multicolumn{2}{|l}{\citet{Jain1992,Straub1996},} \\
8 & $\text{H}_2 + e^- \rightarrow 2\text{H}^+ + 2e^- + e^-$ & \multicolumn{2}{|l}{\citet{Liu1998,Brunger2002}} \\
9 & $\text{H} + e^- \rightarrow \text{H}^+ + e^- + e^-$ & \multicolumn{2}{l}{\citet{Brackmann1958,Burke1962},} \\
  &														& \multicolumn{2}{l}{\citet{Bray1991,MAYOL1997},} \\
  &														& \multicolumn{2}{l}{\citet{Stone2002,Bartlett2004}} \\
10 & $\text{He} + e^- \rightarrow \text{He}^+ + e^- + e^-$ & \multicolumn{2}{l}{\citet{LaBahn1970,MAYOL1997},} \\
   &													   & \multicolumn{2}{l}{\citet{Stone2002,Bartlett2004}} \\
    & & & \\
 & Charge exchange & & \\
 & or proton transfer: & & \\
11 & $\text{H}^+ + \text{H}_2(\nu \geq 4) \rightarrow \text{H}_2^+ + \text{H}$ & \text{see text, Sect.~\ref{sec:ionomodel}} &  \\
12 & $\text{H}_2^+ + \text{H}_2 \rightarrow \text{H}_3^+ + \text{H}$ & $2.0\times10^{-9}$ & \citet{Theard1974} \\
13 & $\text{H}^+ + \text{H}_2 + \text{M} \rightarrow \text{H}_3^+ + \text{M}$ & $3.2\times10^{-29}$ & \citet{Kim1994} \\
14 & $\text{He}^+ + \text{H}_2 \rightarrow \text{H}^+ + \text{H} + \text{He}$ & $1.0\times10^{-9}\text{exp}(-5.7\times10^3/T)$ & \citet{Moses2000} \\
15 & $\text{He}^+ + \text{H}_2 \rightarrow \text{H}_2^+ +\text{He}$ & $9.35\times10^{-15}$ & \citet{Anicich1993} \\
16 & $\text{H}_3^+ + \text{H} \rightarrow \text{H}_2^+ +\text{H}_2$ & $2.1\times10^{-9}\text{exp}(-2.0\times10^4/T)$ & \citet{Harada2010} \\
17 & $\text{H}_2^+ + \text{H} \rightarrow \text{H}^+ +\text{H}_2$ & $6.4\times10^{-10}$ & \citet{Karpas1979} \\
 & & & \\
 & Recombination: & & \\
18 & $\text{H}^+ + e^- \rightarrow \text{H} + h\nu$ & $4.0\times10^{-12}(300/T_e)^{0.64}$ & \citet{Storey1995} \\
19 & $\text{H}_2^+ + e^- \rightarrow 2\text{H}$ & $2.3\times10^{-7}(300/T_e)^{0.4}$ & \citet{Auerbach1977} \\
20 & $\text{He}^+ + e^- \rightarrow \text{He} + h\nu$ & $4.6\times10^{-12}(300/T_e)^{0.64}$ & \citet{Storey1995} \\
21 & $\text{H}_3^+ + e^- \rightarrow \text{H}_2 + \text{H}$ & $2.9\times10^{-8}(300/T_e)^{0.64}$ & \citet{Sundstrom1994} \\
22 & $\text{H}_3^+ + e^- \rightarrow 3\text{H}$ & $8.6\times10^{-8}(300/T_e)^{0.64}$ & \citet{Datz1995} \\
\bottomrule
\end{tabular}
\label{tab:reactionsiono}
\end{table*}

\subsection{H$_3^+$ emission model}
H$_3^+$ emissions from EGP ionospheres may be detectable in optimal circumstances with current and in particular with future ground-based telescopes \citep[e.g.][]{Shkolnik2006}. This would be very interesting because, in addition to providing constraints on the ionosphere, they could be used to detect non-transiting planets. Therefore, we determined emission strengths for the various planets that we modelled (see Sect.~\ref{sec:H3Pemissions}). The H$_3^+$ ion has two vibrational modes, of which only the second, $\nu_2$, is active. It is centred on a wavelength of 3.9662~$\mu$m. The energy emitted in a given line under local thermodynamic equilibrium (LTE) was calculated using the following equation\citep{Miller2010}:
\begin{equation}
I(\omega_{if},T) = \frac{1}{4\pi Q(T)}g_f (2J_f+1) hc\,\omega_{if} A_{if} \text{exp}\left(-\frac{hcE_f}{kT}\right),
\end{equation}
where the indices $i$ indicate the lower state and $f$, the upper state, $\omega_{if}$ is the frequency of the transition (per cm), $A_{if}$ is the Einstein $A$ coefficient of the transition, $g_f$ is the nuclear spin degeneracy, $J_f$ is the angular momentum of the upper state $f$, $E_f$ is the energy of the upper state (per cm), $Q(T)$ is the partition function of H$_3^+$, $h$ is the Planck constant, $c$ is the speed of light, $k$ is the Boltzmann constant, and $T$ is the temperature. The factor $hc$ converts wavenumbers to SI units, meaning that $I$ has units of $\text{W}\,\text{molecule}^{-1}\,\text{sr}^{-1}$. Energy levels and Einstein $A$ coefficients were obtained from \citet{Dinelli1992}. The power emitted in a given line was obtained by multiplying $I$ by the number of H$_3^+$ molecules and considering emission over $2\pi$~sr. Partition functions are those described in \citet{Miller2010}, which are valid for high-temperature atmospheres. 

H$_3^+$ levels are often sub-thermally populated in planetary atmospheres. \citet{Miller2013} estimated that this is the case up to densities of $10^{11} - 10^{12}$~cm$^{-3}$, and that it is due to high Einstein coefficients, meaning that there is competition between radiative relaxation and collisional de-excitation at high densities. The densities we predict are lower than $10^{11}$~cm$^{-3}$ (see Sects.~\ref{sec:stable}, \ref{sec:hydrodyn}, and \ref{sec:vary_a}), it is therefore important to consider non-LTE effects when determining H$_3^+$ emission strengths. These effects were taken into account by introducing a weighting factor, using the method of \citet{Oka2004}, as detailed in \citet{Miller2013}. This weighting factor is dependent on the density of H$_2$.

We determined emission in the Q and R branches, which show stronger emission than the P branch, and we also provide values for the total H$_3^+$ output power. The total output power per H$_3^+$ molecule can be expressed as:
\begin{equation}
E(\text{H}_3^+,T) = \sum_{if} I(\omega_{if},T).
\end{equation}
To calculate this quantity, we used the following parametrisation, determined by \citet{Miller2013}:
\begin{equation}
\text{log}_e \left(E(\text{H}_3^+,T)\right) = \sum_n C_n T^n,
\end{equation}
where the coefficients $C_n$ are provided in Table~5 of \citet{Miller2013}.

\section{Ionisation in EGPs orbiting stars of different activity levels} \label{sec:results}
\subsection{Effect of the photo-absorption cross-section resolution} \label{sec:sig_res}

The photo-absorption cross section of H$_2$ is very structured beyond the ionisation threshold ($\lambda_{\text{th,H}_2}=80.4$~nm), as a result of absorption in the Lyman, Werner, and Rydberg bands. However, all EGP and most solar system ionospheric studies \citep[e.g.][]{Galand2009,Koskinen2010} have previously used the low-resolution measurements of \citet{Backx1976}, taken at 3 -- 4~nm intervals. We evaluated the impact on EGP ionospheres of using a high-resolution H$_2$ photo-absorption cross section with $\Delta\lambda=10^{-3}$~nm (plotted in grey in Fig.~\ref{fig:cross-sec}(a)), instead of the low-resolution \citet{Backx1976} measurements (plotted as a thick black line in Fig.~\ref{fig:cross-sec}(a)). The high-resolution cross section we used consists of the \citet{Backx1976} measurements below $\lambda_{\text{th,H}_2}=80.4$~nm and high-resolution calculations by \citet{Yelle1993} at longer wavelengths. In addition, the H$_2$ photo-dissociation cross section measured by \citet{Dalgarno1969} was added to the \citet{Yelle1993} calculations for wavelengths between 80.4~nm and 84.2~nm. Until now, these cross sections with a high spectral resolution have not been used in studies of EGP ionospheres. However, at Saturn \citep{Kim2014} and Jupiter \citep{Kim1994}, using high-resolution H$_2$ photo-absorption cross sections led to the prediction of a larger and more extended layer of hydrocarbon ions in the lower ionosphere.

We find that high-resolution cross sections are not required when modelling pure H/H$_2$/He EGP atmospheres. The ratio of atomic hydrogen to molecular hydrogen column densities is significantly larger in the extended EGP atmosphere than at Saturn or Jupiter. This means that absorption by H dominates absorption by H$_2$, and as such, the resolution of $\sigma_{\text{H}_2}^{\text{abs}}$ has a much weaker effect for EGPs than for the solar system cases described by \citet{Kim1994} and \citet{Kim2014}. The differences between using high- and low-resolution H$_2$ photo-absorption cross sections are minor (change lower than 2.5~\% in ion densities), therefore we consider that the low-resolution cross sections are sufficient for the remainder of this work on EGP ionospheres.

Even though high-resolution cross sections are not required here, this may not always be the case. If photo-ionisation reactions are included that involve ionisation threshold wavelengths larger than that of H ($\lambda_{\text{th,H}}=91.2$~nm), such as those involving certain hydrocarbons, then the effects of absorption by H$_2$ in the Lyman, Werner and Rydberg bands may no longer be weak. At these wavelengths, there is no longer any absorption by atomic H, and accordingly, absorption over a wide range of pressures by molecular H$_2$ occurs. The inclusion of the high-resolution cross section may therefore allow for more efficient ionisation below the thermosphere.

\subsection{Production rates: primary and secondary ionisation} \label{sec:prodrates}
We included ionisation through both primary (photo-) ionisation and secondary (electron-impact) ionisation processes (see Sect.~\ref{sec:ionomodel}). Electron-impact ionisation operates through collisions with photo-electrons and their secondaries. In this section, we describe the contribution of each of these processes to ionospheric densities. We chose to run the model for EGPs orbiting at 1~AU from the Sun (at minimum and maximum activity), $\epsilon$ Eri, AD Leo, and AU Mic, to study the contribution of the different stellar spectral energy distributions. At this orbital distance, our assumption of a pure H$_2$/H/He thermosphere is valid, as long as the eddy diffusion coefficient is lower than $\sim10^2$ -- $10^3$~m$^2$~s$^{-1}$. Below (Sect.~\ref{sec:vary_a}), we examine the effect of varying the orbital distance.

\begin{figure*}
\sidecaption
\includegraphics[width=12cm]{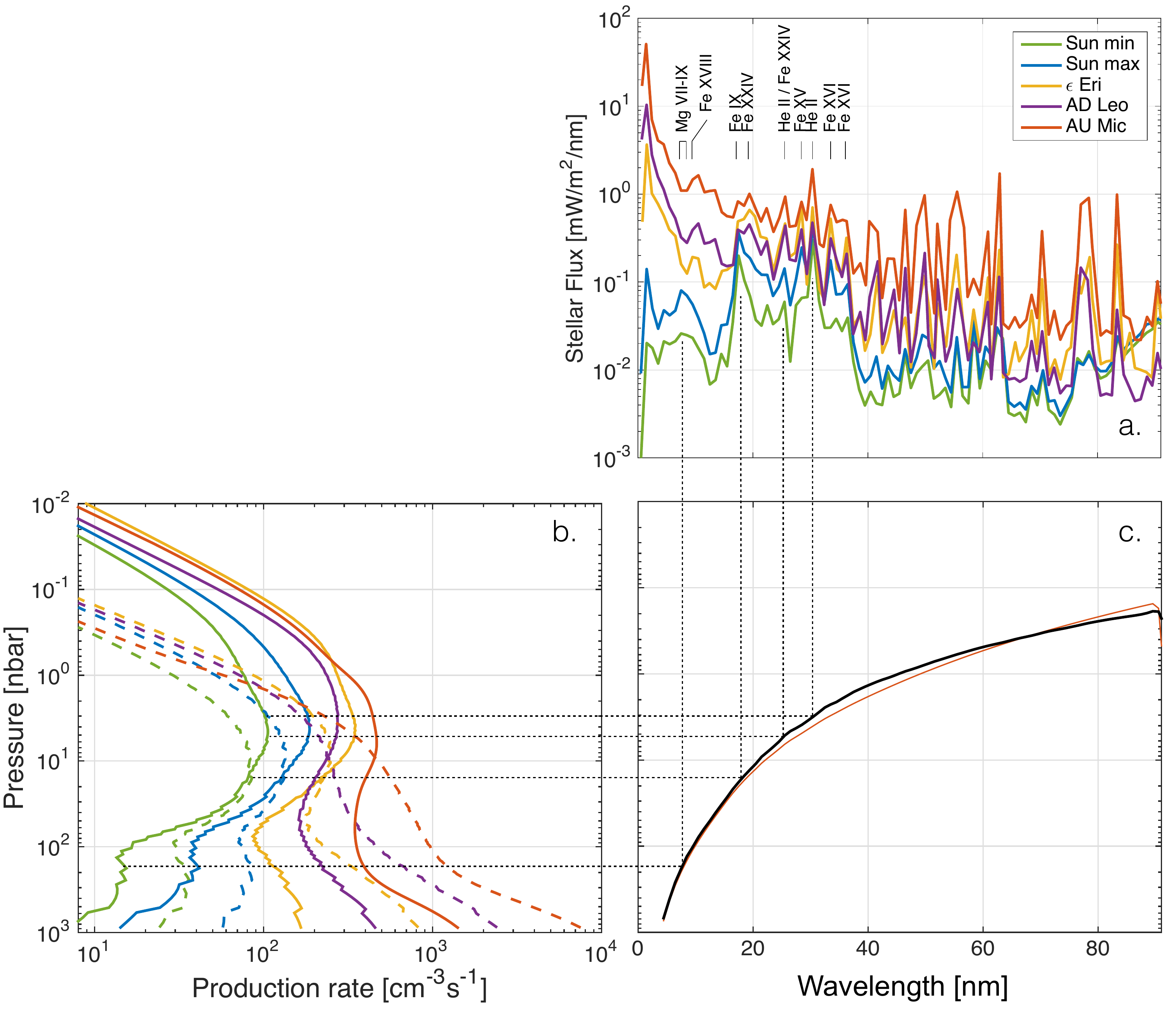}
\caption{(a) Stellar fluxes at 1~AU as a function of wavelength for the Sun at solar minimum (TIMED/SEE observation from 15 May 2008) in green, the Sun at solar maximum (TIMED/SEE observation from 14 January 2013) in blue, $\epsilon$ Eri (synthetic spectrum) in yellow, AD Leo (synthetic spectrum) in red, and AU Mic (synthetic spectrum) in purple. Some major lines that affect the ionospheric peak are labelled. (b) Electron production rate at local noon through photo-ionisation (solid lines) and electron-impact ionisation (dashed lines) for an EPG orbiting at 1 AU from the Sun at solar minimum (green), the Sun at solar maximum (blue), $\epsilon$ Eri (yellow), AD Leo (purple), and AU Mic (red). (c) Pressure for which the optical depth $\tau$ is unity at local noon, as a function of wavelength for planets orbiting the Sun, $\epsilon$ Eri, and AD Leo (think, black line) and for a planet orbiting at 1~AU from AU Mic (thin, red line).}
\label{fig:stars_Pe_ptau1_spectra}
\end{figure*}

Figure~\ref{fig:stars_Pe_ptau1_spectra}(b) shows the total ion production rates for photo-ionisation (solid lines) and electron-impact reactions (dashed lines) for planets orbiting each star. We also show in this figure the stellar spectra used to drive the ionospheric model in panel (a) and the pressures of unity optical depth in panel (c). The pressures $p$ at which $\tau=1$, shown in panel (c), allow us to identify the emission lines in the stellar spectrum (panel (a)) that are the main contributors to the different peaks in the ionisation rate plots (panels (b)). Solar spectra were obtained from TIMED/SEE on 15 May 2008 (F10.7 = 70~solar flux units)  for the case of solar minimum and on 14 January 2013 (F10.7 = 154~solar flux units) for solar maximum. Synthetic spectra from the coronal model were used for the other stars. We note that $p(\tau=1)$ is dependent on the neutral atmosphere, but since the neutral atmospheres of planets in the same escape regime are so similar, the curves of $p(\tau=1)$ are also very similar. We have plotted the mean of the four $p(\tau=1)$ curves for planets in hydrostatic equilibrium as a thick black line in panel (c): those of planets around the Sun at minimum and maximum, $\epsilon$ Eri and AD Leo. Had we plotted each case individually, the difference would barely have been noticeable because the curves are mostly superposed. The curve of $p(\tau=1)$ for the case of a planet orbiting AU Mic is slightly different from the others (plotted as a light red line in panel (c)), since this planet is in a regime of hydrodynamic escape (see Sects.~\ref{sec:stable} and \ref{sec:hydrodyn}).

The optical depth is determined by the densities of atmospheric constituents, the incoming stellar flux, and the photo-absorption cross sections (plotted in Fig.~\ref{fig:cross-sec}). Since we used low-resolution photo-absorption cross sections (see Sect.~\ref{sec:sig_res}), the pressure of unity optical depth decreases monotonically with photon wavelength (i.e. the altitude of unity optical depth increases monotonically with wavelength). This means that higher energy photons deposit their energy lower in the ionosphere. We also note that due to the higher gradient of the $p(\tau=1)$ curve at lower wavelengths, which is associated with a large change in $\sigma^{\text{abs}}$, high-energy (e.g. soft X-ray, $\sim 0.1 - 10$~nm) stellar emission lines affect a much more extended region of atmosphere than higher wavelength stellar emissions.

Because the spectral shapes are similar, the ionisation rate profiles of planets orbiting the Sun at solar min and max are very similar. The main photo-ionisation peak at around 6~nbar is due to the He II lines at 25.6~nm and 30.4~nm in the solar spectrum. These lines are also responsible for the main peaks in the photo-ionisation profiles in planets orbiting $\epsilon$ Eri, AD Leo, and AU Mic. The main peak in electron-impact ionisation in planets irradiated by the solar spectrum occurs at around 10 -- 20~nbar and is formed by the strong Fe emission lines at wavelengths just below 20~nm. There is a secondary peak in the photo-ionisation and electron-impact ionisation profiles of EGPs around a solar-like star, at a pressure of 200 -- 300~nbar. These secondary peaks are due to solar soft X-ray emission just below 10~nm.

The spectral shape of high-energy emissions is quite different in stars that are more active than the Sun. There is a gradual rise in irradiance with decreasing wavelength below about 20~nm that is responsible for an increase in ionisation rates with pressure above about 100~nbar. It seems that the peak ionisation rate is not always reached in the pressure domain of the model, especially for planets around the most active star, AU Mic. So for planets around active stars, there is significant absorption of high-energy radiation and ionisation by associated secondary electrons below 1~$\mu$bar, in the stratosphere. \citet{Koskinen2014a} showed that in close-in EGPs, ionisation of alkali metals, such as Na, K, and Mg, by far-ultraviolet (FUV) photons creates a strong ionisation layer in the stratosphere. Our results here indicate that this ionisation layer in the lower ionosphere can be enhanced by the ionisation of hydrogen by X-rays around active stars.

Another interesting effect of different spectral energy distributions appears in the ionisation profiles of planets around $\epsilon$ Eri and AD Leo. As examined in \citet{Chadney2015}, irradiance in the X-ray and EUV wavebands evolves at different rates over the lifetime of the star \citep[see also][]{Ayres1997,Ribas2005,Sanz-Forcada2011}. High-energy stellar emissions decrease as the star ages, but the highest energy emissions decrease at a faster rate. Indeed, \citet{Chadney2015} showed that the ratio of fluxes at the stellar surface, $F_{\text{EUV}}/F_{\text{X}}$ increases as the star ages. $\epsilon$ Eri is an older star than AD Leo and so has lower emissions in the X-ray, but still has higher emissions in the EUV. This results in higher ionisation in the lower part of the ionosphere in a planet orbiting AD Leo than $\epsilon$ Eri, but lower ionisation at higher altitudes. We note that this is the case even though AD Leo is smaller in size than $\epsilon$ Eri, based on which lower emission might have been expected for the same activity level. AD Leo is much younger and hence more active, however, which is a stronger factor than size.

\begin{figure}
\centering
\includegraphics[width=0.48\textwidth]{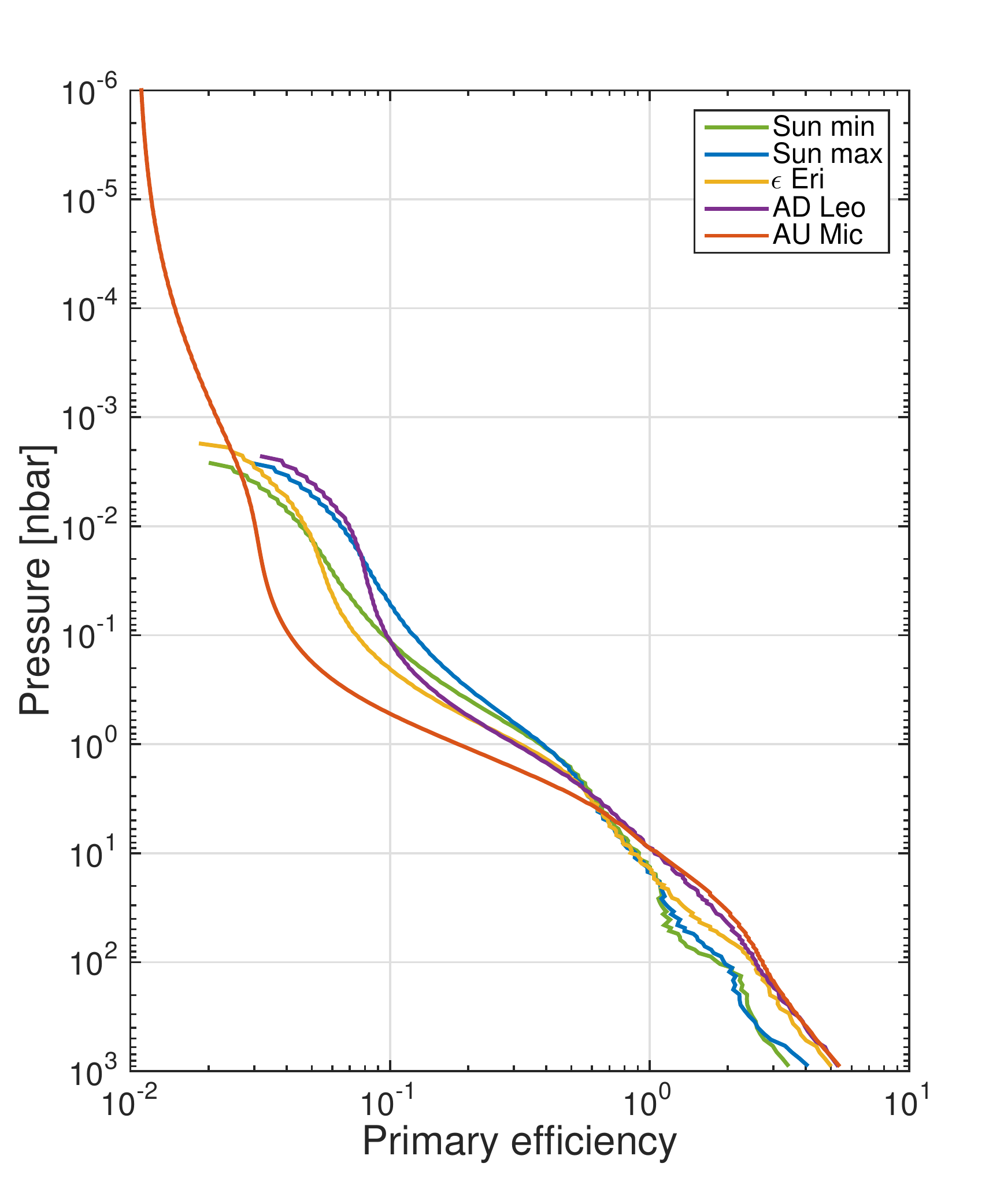}
\caption{Primary efficiency (ratio of secondary to primary electron production rate) versus pressure in an EGP orbiting at 1~AU from the Sun at solar minimum (green), the Sun at solar maximum (blue), the star $\epsilon$ Eri (yellow), AD Leo (purple), and AU Mic (red).}
\label{fig:primary_efficiency_stars}
\end{figure}

The primary efficiency is defined as the ratio of electron-impact ionisation rate to photo-ionisation rate. This is plotted in Fig.~\ref{fig:primary_efficiency_stars} for each of the five cases. As expected, since the pressure at which $\tau=1$ increases monotonically with decreasing wavelength (see Fig.~\ref{fig:stars_Pe_ptau1_spectra}(c)), higher energy photons are absorbed lower in the ionosphere. This means that the highest energy photo-electrons, which are responsible for most of the ionisation, are formed at high pressures. The pressure at which electron-impact ionisation overtakes photo-ionisation is about 10~nbar, which is just below the main production peak (see Fig.~\ref{fig:stars_Pe_ptau1_spectra}(b)). Thus electron-impact ionisation increases the electron densities significantly below the main EUV ionisation peak, allowing for the lower atmosphere to be ionised more efficiently than in models that ignore electron-impact ionisation.

\subsection{Ionospheric densities: atmosphere in hydrostatic equilibrium} \label{sec:stable}
\begin{figure*}
	\sidecaption
	\includegraphics[width=12cm]{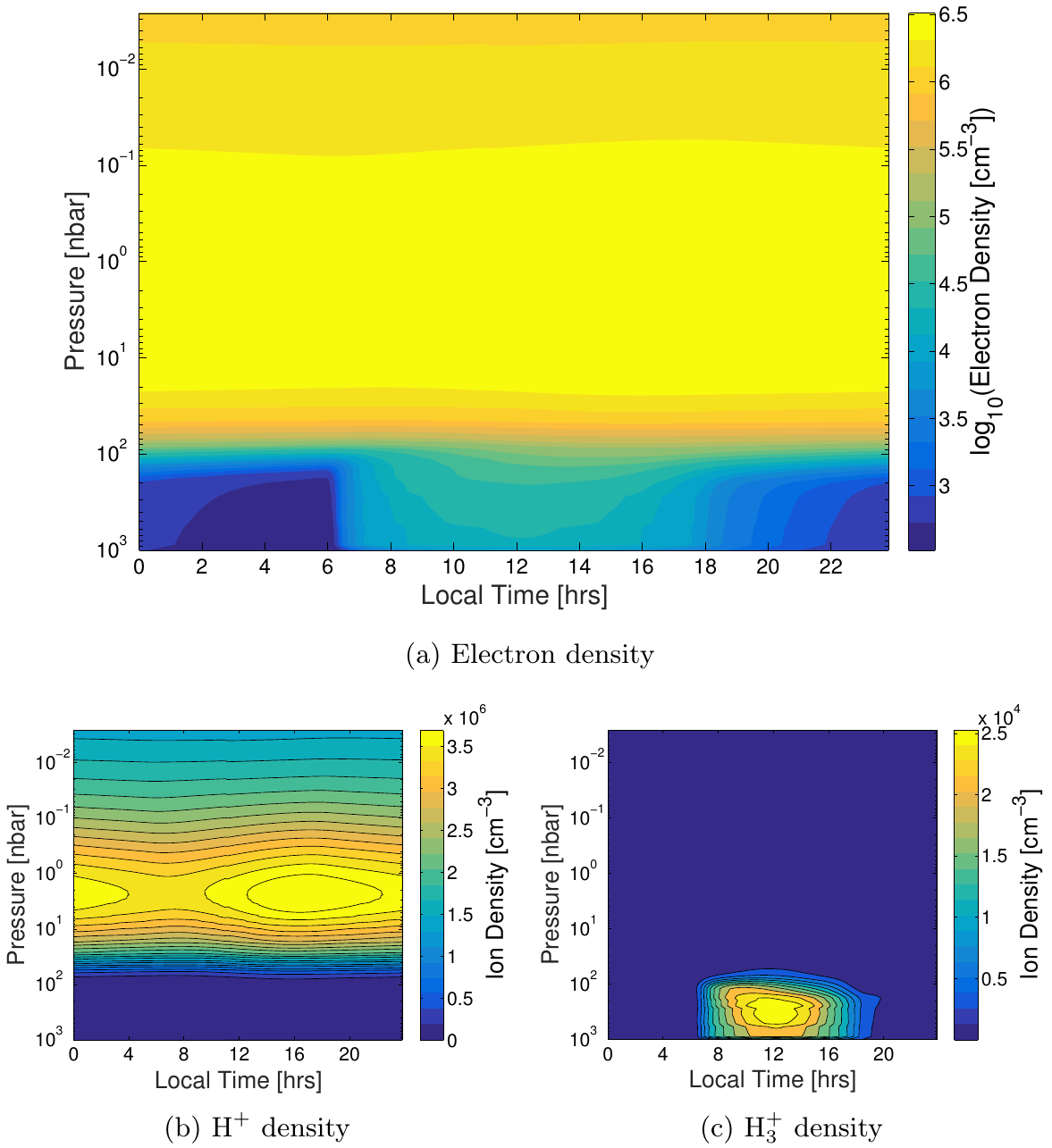}
	\caption{Densities as a function of local time and pressure for an EGP at 1~AU from a Sun-like star (at solar maximum), shown at 30$^{\circ}$ latitude. The electron densities in plot (a) are shown on a log scale, whereas the ion densities in plot (b) and (c) are plotted according to a linear scale.}
\label{fig:densiy_contourplot}
\end{figure*}

This section describes the ionospheric densities in planets with atmospheres in hydrostatic equilibrium. These are planets orbiting at larger distances from their stars, with atmospheres that undergo thermal Jeans escape. For the cases studied here, EGPs around the Sun, $\epsilon$ Eri, and AD Leo, at an orbital distance of 1~AU, are in hydrostatic equilibrium. EGPs at 1~AU from the more active star AU Mic have atmospheres that lose material according to an organised outflow. The ion densities in these atmospheres, which are in the hydrodynamic escape regime, are detailed in Sect.~\ref{sec:hydrodyn}.

The variation of electron density, as well as $n_{\text{H}^+}$ and $n_{\text{H}_3^+}$ over the course of a day, is presented in the contour plots in Fig.~\ref{fig:densiy_contourplot}, at a latitude of 30$^{\circ}$ for a planet orbiting the Sun at 1~AU. We note that the electron densities in Fig.~\ref{fig:densiy_contourplot}(a) are plotted on a logarithmic scale, whereas the ion densities of H$^+$ (Fig.~\ref{fig:densiy_contourplot}(b)) and H$_3^+$ (Fig.~\ref{fig:densiy_contourplot}(c)), are plotted on a linear scale. This is so that the two distinct regions are clearly visible in the electron density plot. They correspond to an $\text{H}^+$-dominated region in the upper ionosphere at $p\lesssim200$~nbar, and to an $\text{H}_3^+$-dominated region below. The $\text{H}^+$-dominated region is relatively uniform over the diurnal cycle, with a peak spread out over a wide pressure range.

The $\text{H}_3^+$-dominated region at $p\gtrsim200$~nbar clearly displays a strong diurnal variation (see also Fig.~\ref{fig:densiy_contourplot}(c)), with a density peak near 12 LT. H$_3^+$ densities decrease gradually in the evening, as photo-ionisation of H$_2$ diminishes with the setting star and the remaining $\text{H}_3^+$ is destroyed through electron dissociative recombination reactions 21 and 22. At dawn, as photo-ionisation is switched back on, there is a more rapid build-up of $\text{H}_3^+$ ions.

The diurnal variations in the H$^+$-dominated region are less pronounced, but can be seen in Fig.~\ref{fig:densiy_contourplot}(b) that shows H$^+$ densities. Unlike the $\text{H}_3^+$ density peak, the H$^+$ peak is offset with the stellar illumination peak. A maximum H$^+$ density of $3.9\times 10^6$~cm$^{-3}$ occurs at about 17 LT, the smallest peak is at 7 LT, where $n_{\text{H}^+}=3.4\times 10^6$~cm$^{-3}$. This lag behind the stellar illumination is due to the fact that H$^+$ is a long-lived ion. 

\begin{figure}
\centering
\includegraphics[width=0.49\textwidth]{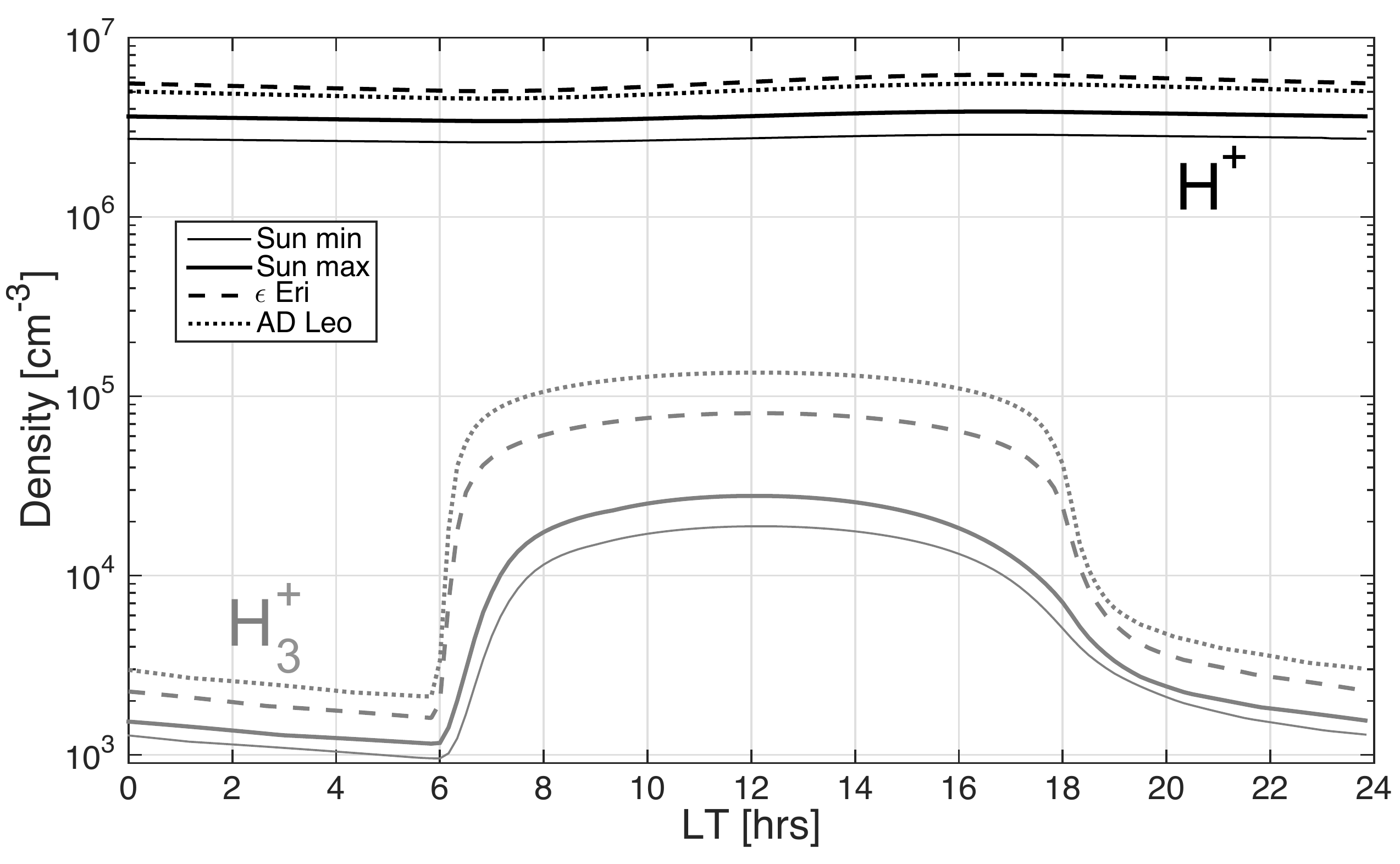}
\caption{Diurnal variation of peak H$^+$ (in black) and H$_3^+$ (in grey) densities in an EGP orbiting the Sun at solar minimum (thin solid lines), the Sun at solar maximum (thick solid lines), the star $\epsilon$ Eri (dashed lines), and the star AD Leo (dotted lines).}
\label{fig:peak_nHP_nH3P_vs_LT_stars}
\end{figure}

Figure~\ref{fig:peak_nHP_nH3P_vs_LT_stars} presents the variation over the course of a day of the peak densities and timings of the two main ionospheric constituents, H$^+$ (in black) and H$_3^+$ (in grey), for planets with atmospheres in hydrostatic equilibrium, orbiting each star. The same diurnal variation is predicted as for a planet orbiting the Sun during solar maximum, namely a strong variation in densities of the short-lived ion H$_3^+$, with a peak that is almost symmetrical around 12 LT, and a significantly smaller, almost imperceptible diurnal variation in H$^+$ density, with a peak at around 17 LT that is offset from the peak in stellar illumination. Peak H$_3^+$ densities increase along with the stellar X-ray flux, whereas peak H$^+$ densities are a function of the stellar EUV flux level. Thus, higher H$_3^+$ densities occur in the planet around AD Leo than around $\epsilon$ Eri, but higher H$^+$ densities are found in the planet orbiting $\epsilon$ Eri. Values of peak H$_3^+$ and H$^+$ densities are provided in Table~\ref{tab:peak_ni_1au}.

\begin{figure}
\centering
\includegraphics[width=0.48\textwidth]{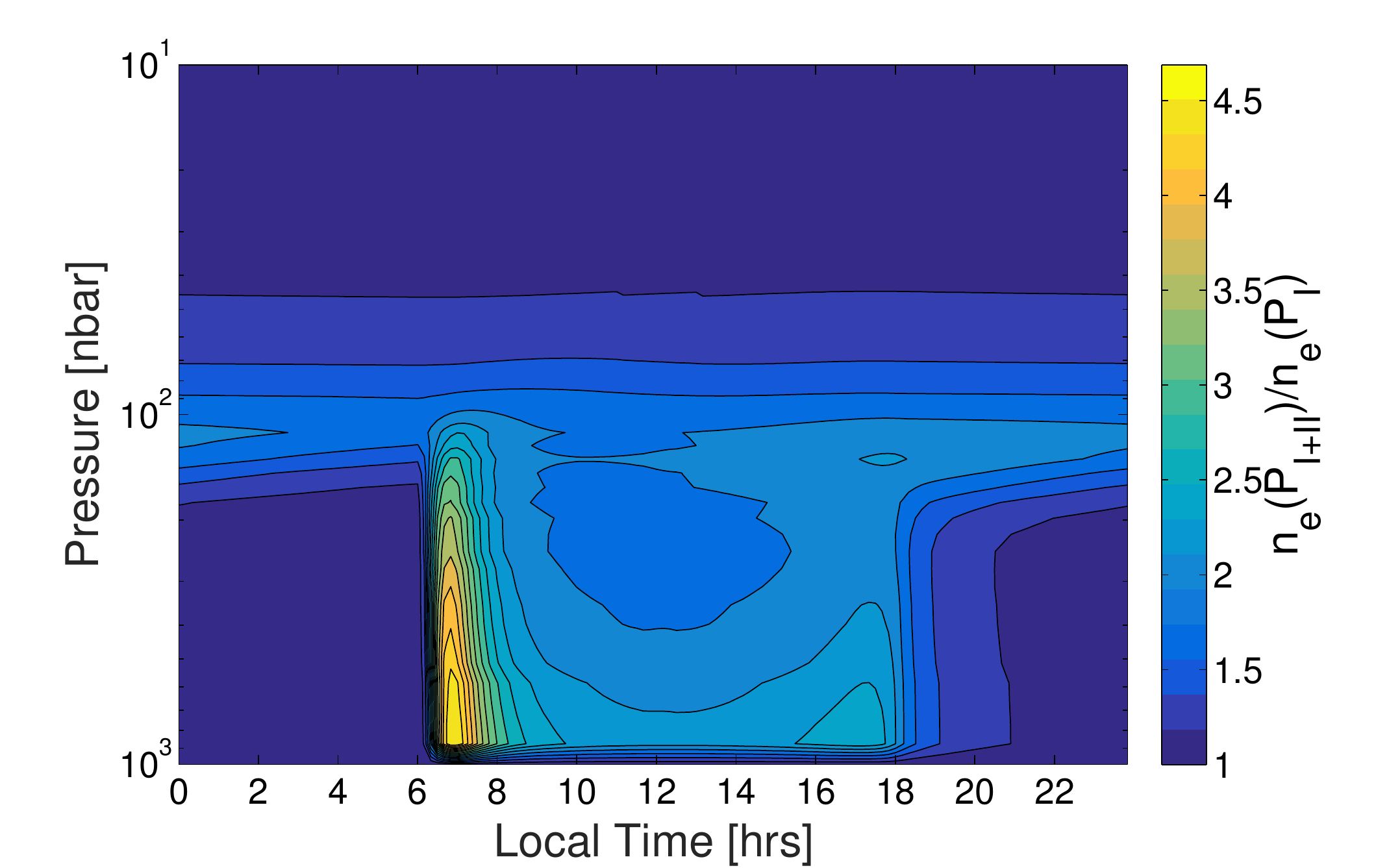}
\caption{Ratio of electron densities determined including both primary (photo-) ionisation and secondary (electron-impact) ionisation $n_e(P_{I+II})$ and those determined including just primary ionisation $n_e(P_{I})$, for an EGP orbiting a Sun-like star at 1~AU.}
\label{fig:neIIplusI/neI}
\end{figure}

Electron-impact ionisation affects ion densities in the ionosphere differently at different times of day. Figure~\ref{fig:neIIplusI/neI} is a contour plot showing the ratio of electron densities between a full calculation taking both photo- and electron-impact ionisation into account ($n_e(P_{I+II})$) and electron densities calculated with only photo-ionisation ($n_e(P_{I})$). These are shown for an EGP orbiting a Sun-like star at 1~AU, but are qualitatively similar for EGPs orbiting the other stars. Only the lower ionosphere is plotted since above a few 10$^1$~nbar, essentially $n_e(P_{I+II})/n_e(P_{I})\sim 1$. Including secondary ionisation has the strongest effect on electron densities shortly after dawn, at around 7 LT, which corresponds to the local time when H$_3^+$ densities increase strongly. This strong increase in the ratio just after dawn occurs because the strong increase in electron-impact ionisation after sunrise is combined with a low loss rate for the ionospheric plasma below $10^2$~nbar. In this region H$_2^+$ is the main ion produced under stellar illumination and is rapidly converted to H$_3^+$ (reaction 12). The main loss of H$_3^+$ is through electron dissociative recombination (reactions 21 and 22). After a night without ionisation, the ionosphere is strongly depleted, and therefore the loss rate for the ionospheric plasma is low. In addition, near sunrise the electron-impact ionisation contributes up to $10^2$~nbar, while at noon it is confined towards lower altitudes (higher pressures). The reason is that photo-electrons that are energetic enough to cause significant ionisation are produced by solar photons with wavelength shorter than 20~nm. Figure~\ref{fig:stars_Pe_ptau1_spectra}(c) shows that these photons are absorbed primarily below $10^1$~nbar at noon. At larger stellar zenith angles, the altitude of deposition increases, which explains the extension towards lower pressures of the contribution of electron-impact ionisation at sunrise. Since there are no photo-electrons during the night, there is little electron-impact ionisation. There are some differences after dusk, while extra ions are destroyed that were built up during the day though secondary ionisation.

\begin{table*}
\centering
\caption{Peak densities of H$^+$ and H$_3^+$ over a diurnal cycle, at 30$^{\circ}$ latitude in the ionosphere of planets orbiting different stars at a distance of 1~AU.}
\begin{tabular}{l|c|c}
\toprule
 & max($n_{\text{H}_3^+}$) [cm$^{-3}$] &  max($n_{\text{H}^+}$) [cm$^{-3}$] \\
\midrule
Sun at solar min & $1.9\times10^4$ &  $2.9\times10^6$ \\
Sun at solar max & $2.8\times10^4$ &  $3.9\times10^6$ \\
$\epsilon$ Eri & $8.0\times10^4$ &  $6.2\times10^6$ \\
AD Leo & $1.4\times10^5$ & $5.5\times10^6$ \\
\bottomrule
\end{tabular}
\label{tab:peak_ni_1au}
\end{table*}

\subsection{Ionospheric densities: atmosphere in hydrodynamic escape regime} \label{sec:hydrodyn}
\begin{figure*}
	\sidecaption
	\includegraphics[width=12cm]{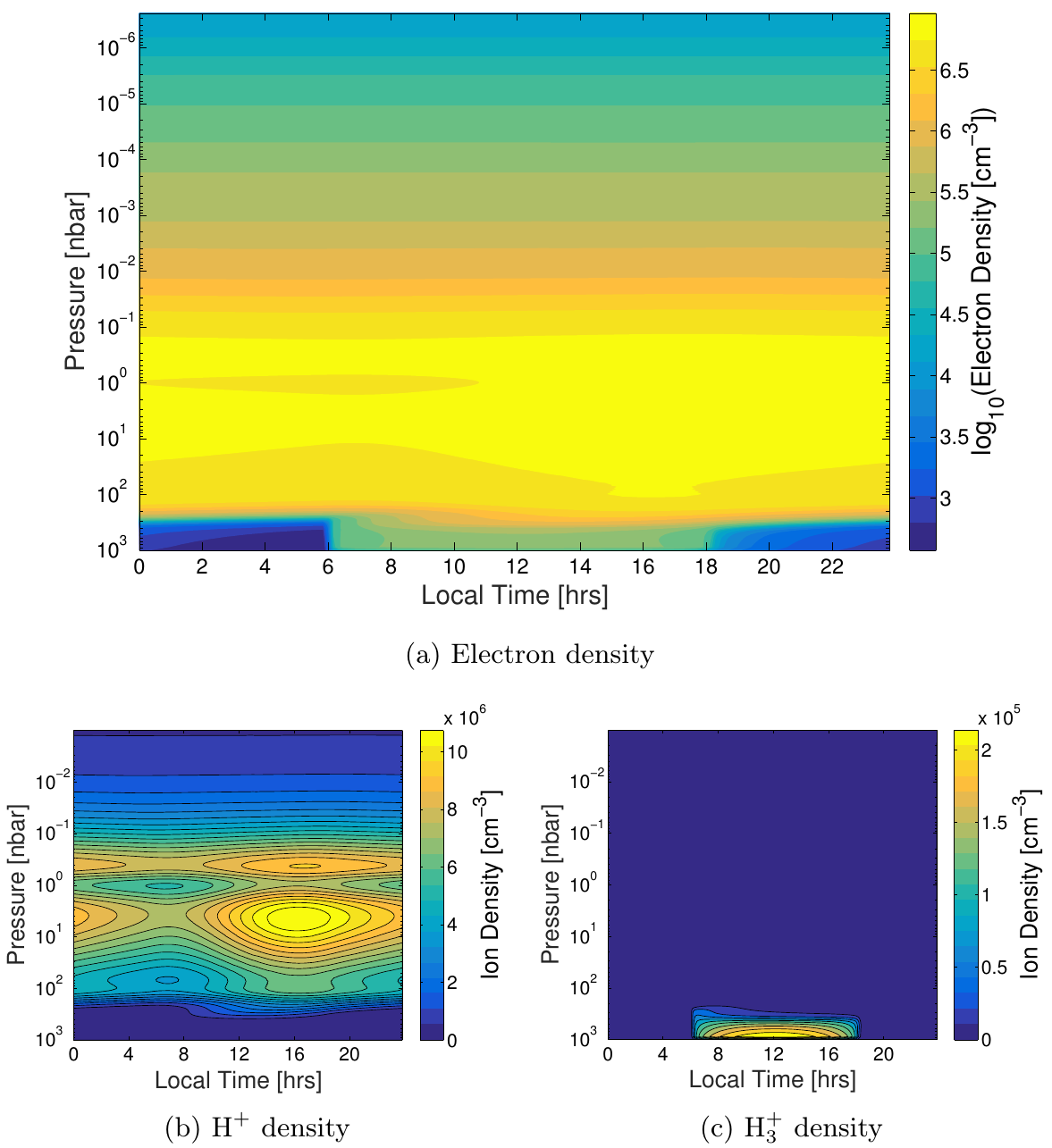}
\caption{Densities as a function of local time and pressure, for an EGP at 1~AU from the star AU Mic, shown at 30$^{\circ}$ latitude. Note that the electron densities in plot (a) are shown on a log scale, whereas the ion densities in plot (b) and (c) are plotted according to a linear scale.}
\label{fig:densiy_contourplot_aumic1au}
\end{figure*}

AU Mic is the youngest and most active star included in this study, and as such has the strongest XUV emissions. As a consequence, the HD209458b-type planet that we modelled is in a regime of hydrodynamic escape, even at an orbital distance of 1~AU, whereas the same planet orbiting at the same distance from a Sun-like star, $\epsilon$ Eri, or AD Leo is in the Jeans escape regime \citep{Chadney2015}. The atmosphere is more extended than at planets in the stable regime and molecules, such as H$_2$, are fully dissociated at high pressures. In the case of an EGP orbiting AU Mic at 1~AU, H$_2$ is dissociated between 1 and 0.1~nbar.

Similarly as in planets in the stable regime, the ionosphere of EGPs undergoing hydrodynamic escape is composed of two distinct regions, characterised by the major ion: H$^+$ is dominant above the H$_2$/H dissociation front, whereas H$_3^+$ dominates below. H$_2^+$ and He$^+$ ions are always minor constituents.

Contour plots of the electron and major ion densities are shown in Fig.~\ref{fig:densiy_contourplot_aumic1au} at 30$^{\circ}$ latitude on an EGP orbiting AU Mic at 1~AU. As in the planets orbiting less active stars, H$_3^+$ ions reach a well-defined daily peak near noon and are depleted during night time, when there is no stellar illumination. This is due to their short-lived nature. Conversely, the long-lived H$^+$ ions reach their daily peak density between 16 and 17 LT and are present in large quantities throughout the night because they only recombine slowly with electrons (reaction 18 is slow).

In contrast to the planets in the stable regime, H$_3^+$ is confined to higher pressures in planets orbiting AU Mic at 1~AU, since it can only be formed in regions where H$_2$ is present. Its peak density is higher than in the planets discussed in Sect.~\ref{sec:stable}, however, as a result of the enhanced X-ray flux: the H$_3^+$ peak density reaches $2.4\times10^5$~cm$^{-3}$ at 12.2 LT (compared to $1.4\times10^5$~cm$^{-3}$ for a planet at 1~AU from AD Leo, see Table~\ref{tab:peak_ni_1au}). Densities of H$^+$ are also higher, a peak of $1.1\times10^7$~cm$^{-3}$ is present at 16.3 LT, at a pressure of about 4~nbar (compared to a peak of $5.5\times10^6$~cm$^{-3}$ at 1~AU from AD Leo). We note that the H$^+$ contour has a double peak, one at 4~nbar and the other at 0.3~nbar (see Fig.\ref{fig:densiy_contourplot_aumic1au}(b)). This is due to the dissociation of H$_2$, which takes place between these two peaks. In particular, reaction 17 ($\text{H}_2^+ + \text{H} \rightarrow \text{H}^+ +\text{H}_2$) only contributes to the peak at 4~nbar. Additionally, below the H$_2$/H dissociation front, photo-ionisation reactions 2 and 3 (ionisation of H$_2$ to form H$^+$) contribute to the peak at 4~nbar. The smaller peak at 0.3~nbar is mainly formed from photo-ionisation of atomic H.

The cross-over mass equation from \citet{Hunten1987}, given in Eq.~3.2 in \citet{Koskinen2014} provides a rough estimate of whether escaping heavy species from the stratosphere can be present in the thermosphere. The expression derived in \citet{Koskinen2014} is independent of altitude for an isothermal atmosphere. We applied this expression here using temperatures that are consistent with our model results. Heavy species, including hydrocarbons and water, could destroy the H$_3^+$ layer. Applying the cross-over mass equation gives a limiting mass-loss rate of $1.5\times 10^5$~kg~s$^{-1}$ required for C to be present at a pressure of 500~nbar (in the H$_3^+$ layer) in an EGP at 1~AU from AU Mic. The upwelling mass flux at this same pressure level is $5.6\times 10^4$~kg~s$^{-1}$, which is thus insufficient to drag heavy species into this region. This gives us confidence that our assumption that species heavier than He are confined below a pressure of 1~$\mu$bar is correct for an EGP at 1~AU from AU Mic.

\subsection{Variations in ion densities with orbital distance} \label{sec:vary_a}
\begin{figure}
\centering
\includegraphics[width=0.49\textwidth]{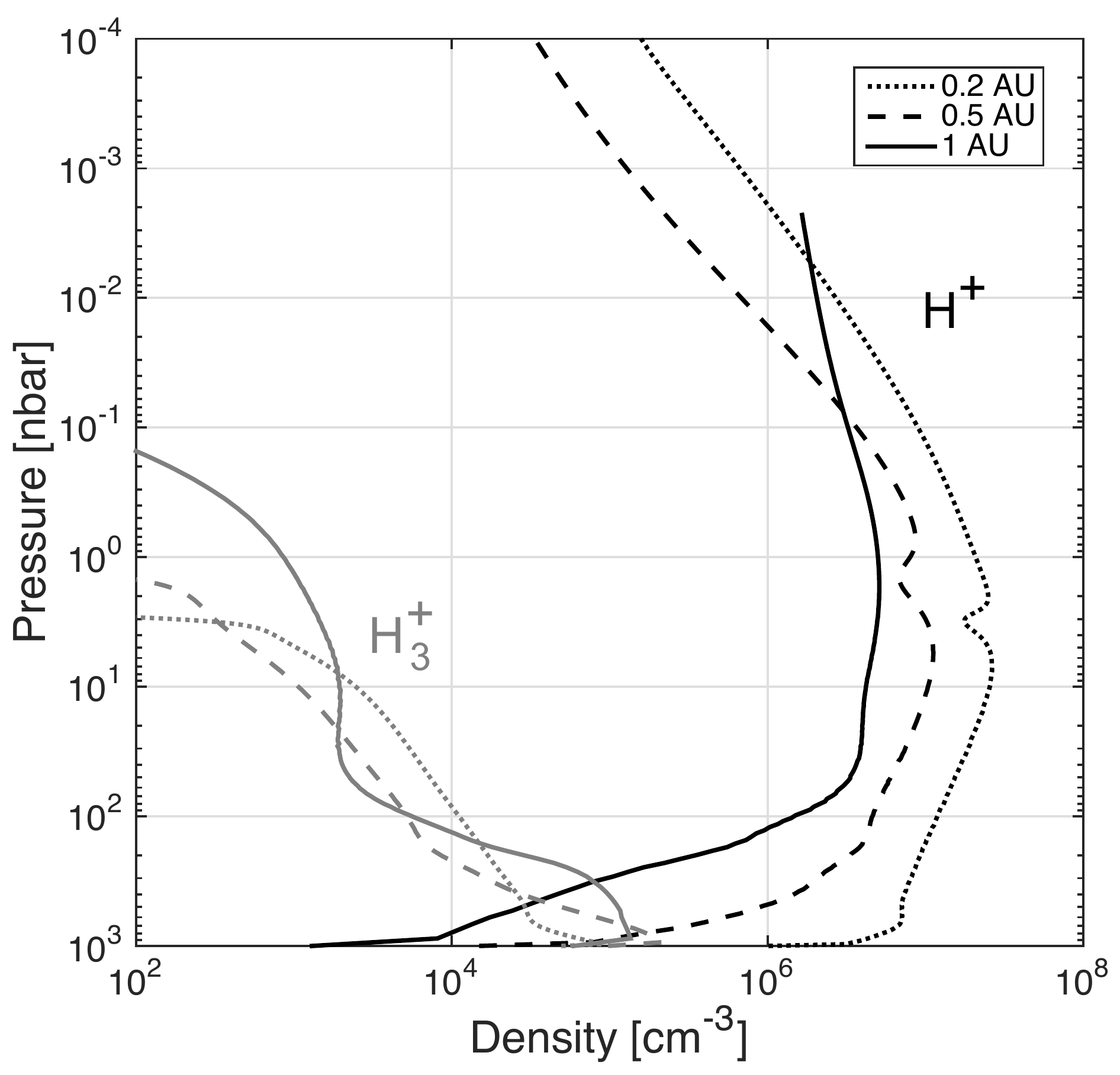}
\caption{Major ion densities for an EGP orbiting AD Leo at different orbital distances, determined at 30$^{\circ}$ latitude and 12 LT.}
\label{fig:ni_vs_p_adleo_diffa}
\end{figure}

In this section, we discuss the changes in the ionosphere with orbital distance by considering an HD209458b-type planet orbiting the star AD Leo at 1~AU, 0.5~AU, and 0.2~AU. The critical orbital distance at which this EGP transitions from Jeans to hydrodynamic escape is located between 1~AU and 0.5~AU \citep{Chadney2015}. Peak temperatures in the thermosphere are about 2,000~K at 1~AU, 8,000~K at 0.5~AU and 10,500~K at 0.2~AU. Ion densities of the two major ions H$^+$ and H$_3^+$, determined at noon at 30$^{\circ}$ latitude, are plotted in Fig.~\ref{fig:ni_vs_p_adleo_diffa}.

As the planet is moved closer to the star, overall ion densities increase with increased stellar radiation, but not in a linear fashion. Within 0.5~AU, the atmosphere escapes hydrodynamically. H$_2$ is fully dissociated below about 1~nbar at 0.5~AU and 3~nbar at 0.2~AU. H$_3^+$ cannot be formed below this dissociation pressure. At small orbital distances, the density of H$^+$ increases vastly in the lower ionosphere: at a pressure of 500~nbar, it increases from $2.7\times10^4$cm$^{-3}$ at 1~AU to $8.9\times10^5$cm$^{-3}$ at 0.5~AU and $7.1\times10^6$cm$^{-3}$ at 0.2~AU. The evolution of the peak density of H$_3^+$ with orbital distance is slightly more complicated. H$_3^+$ is effectively destroyed by electron-recombination (reactions 21 and 22, see Table~\ref{tab:reactionsiono}). Because much more H$^+$ is formed at high pressures, the increase in electron density results in a higher loss rate for H$_3^+$. Therefore, as the H$^+$ peak is pushed to higher pressures with increased stellar flux, the H$_3^+$ peak is reduced. At 0.2~AU, the H$_3^+$ peak is below the lower boundary of the ionospheric model. At higher pressures, the likelihood that H$_3^+$ will further be destroyed through reaction with heavy species is higher. At small orbital distances H$_3^+$ is therefore only likely to be a minor constituent of the ionosphere at all pressure levels.

We obtained a rough estimate of whether escaping heavy species from the stratosphere can be present in the thermosphere by applying the cross-over mass equation from \citet{Hunten1987}. For an EGP at 0.2~AU from AD Leo, the limiting mass-loss rate required for C to be present at a pressure of 100~nbar is $2.7\times 10^6$~kg~s$^{-1}$; and at this same pressure level, the upwelling mass flux is $3.1\times 10^6$~kg~s$^{-1}$. It is thus likely that carbon species are present at this pressure level, which will deplete the H$_3^+$ layer in planets orbiting within 0.2~AU from AD Leo. These results indicate that close-orbiting hot-Jupiter exoplanets are not good candidates for observations of H$_3^+$ emissions. The level of H$_3^+$ IR emissions from the EGP atmospheres modelled in this work is presented in Sect.~\ref{sec:H3Pemissions}.

\subsection{Scaled solar spectra versus synthetic spectra} \label{sec:scaled_spec}
\begin{figure*}
\centering
\includegraphics[width=0.98\textwidth]{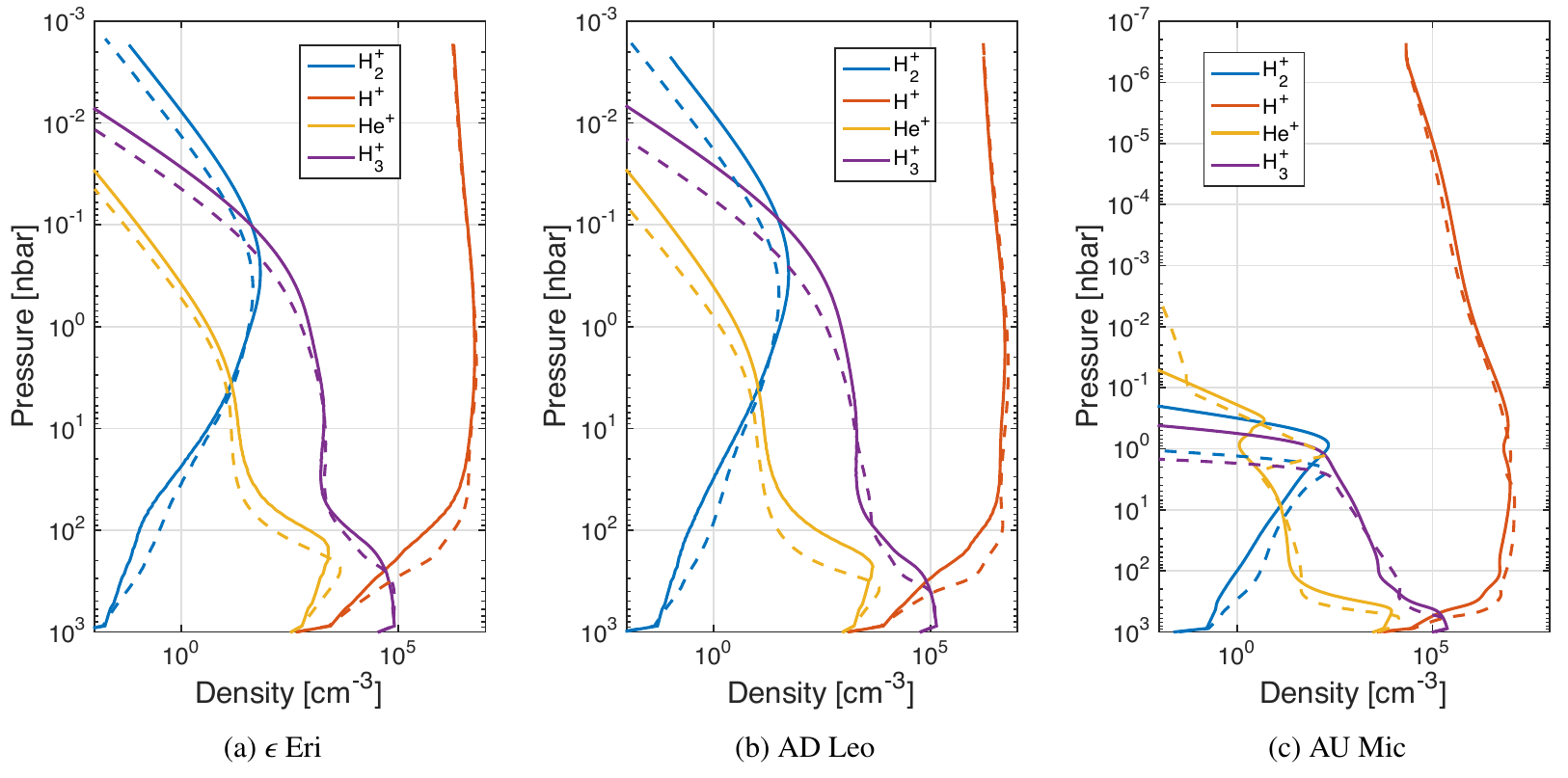}
\caption{Ion densities as a function of pressure at 30$^{\circ}$ latitude and 12 LT for an EGP irradiated by the synthetic spectra (solid lines) and the scaled solar spectra (dashed lines). The orbital distance is 1~AU.}
\label{fig:ni_vs_p_ss}
\end{figure*}

\citet{Chadney2015} derived a new scaling of the solar XUV spectrum to match the spectra of other low-mass stars, using different scaling factors for the X-ray and EUV bandpasses. These scaled spectra were tested against synthetic spectra derived from the coronal model in terms of the neutral atmosphere. It was found that a scaling of the solar spectrum, using two different scaling factors, was adequate to estimate neutral densities and planetary mass-loss rates. In this section, we undertake the same comparison (synthetic to scaled solar spectra) using the ionospheric model. All of the model runs in this section are for planets with an orbital distance of 1~AU. At this orbital distance, the upper atmosphere of EGPs around AU Mic undergo hydrodynamic escape, whereas those orbiting $\epsilon$ Eri and AD Leo are in the classical Jeans escape regime.

Ion densities for planets irradiated by either the synthetic spectra or the scaled solar spectra corresponding to each of the three stars are shown in Fig.~\ref{fig:ni_vs_p_ss}. Ion densities determined using the synthetic spectra are plotted as solid lines and those determined using the scaled spectra are plotted as dashed lines. All of the densities shown in this figure are determined at noon at 30$^{\circ}$ latitude.

At high pressures, more ions are present in the planets irradiated by the scaled spectra. This results in a lowering of the peak in H$^+$ density in planets orbiting all three stars. There is roughly an order of magnitude more H$^+$ at 200~nbar in planets around $\epsilon$ Eri and AD Leo and at 300~nbar in planets around AU Mic. However, at high pressures, densities of H$_3^+$ are either the same or slightly lower when using the scaled spectra. The density of H$_3^+$ is determined by a competition between its production from H$_2^+$ through reaction 12 (see Table~\ref{tab:reactionsiono}) and its loss through electron-recombination (reactions 21 and 22). At high pressures (between about 50 and 500~nbar), there is a significant excess of H$^+$ formed through photo- and electron-impact ionisation with the scaled spectra when compared to using the synthetic spectra. Thus there are also significantly more electrons at these pressures and the increased rates of reactions 21 and 22 counterbalance the increased production of H$_3^+$.

At low pressures, the densities of the minor species H$_3^+$, H$_2^+$, and He$^+$ are higher in the planets irradiated by synthetic spectra than in those irradiated by the scaled spectra because of more photo- and impact-ionisation. Densities of the major ion H$^+$ above about 10~nbar are very similar whether using the synthetic or the scaled solar spectra. 

The difference in terms of ion production and densities is greater between using the scaled solar spectra and the synthetic stellar spectra than in terms of neutral densities. H$^+$ and electron densities vary by about an order of magnitude in the lower ionosphere. However, the density of H$_3^+$ is relatively well regulated and does not vary dramatically at pressures where it is most abundant. This means that at least in the cases studied here, cooling through IR H$_3^+$ emissions is not significantly affected by using scaled solar spectra. Hence the regime of atmospheric escape (and associated mass-loss rates) are well determined using scaled solar spectra with the thermospheric model.

\section{H$_3^+$ emissions} \label{sec:H3Pemissions}
\begin{figure}
\centering
\includegraphics[width=0.49\textwidth]{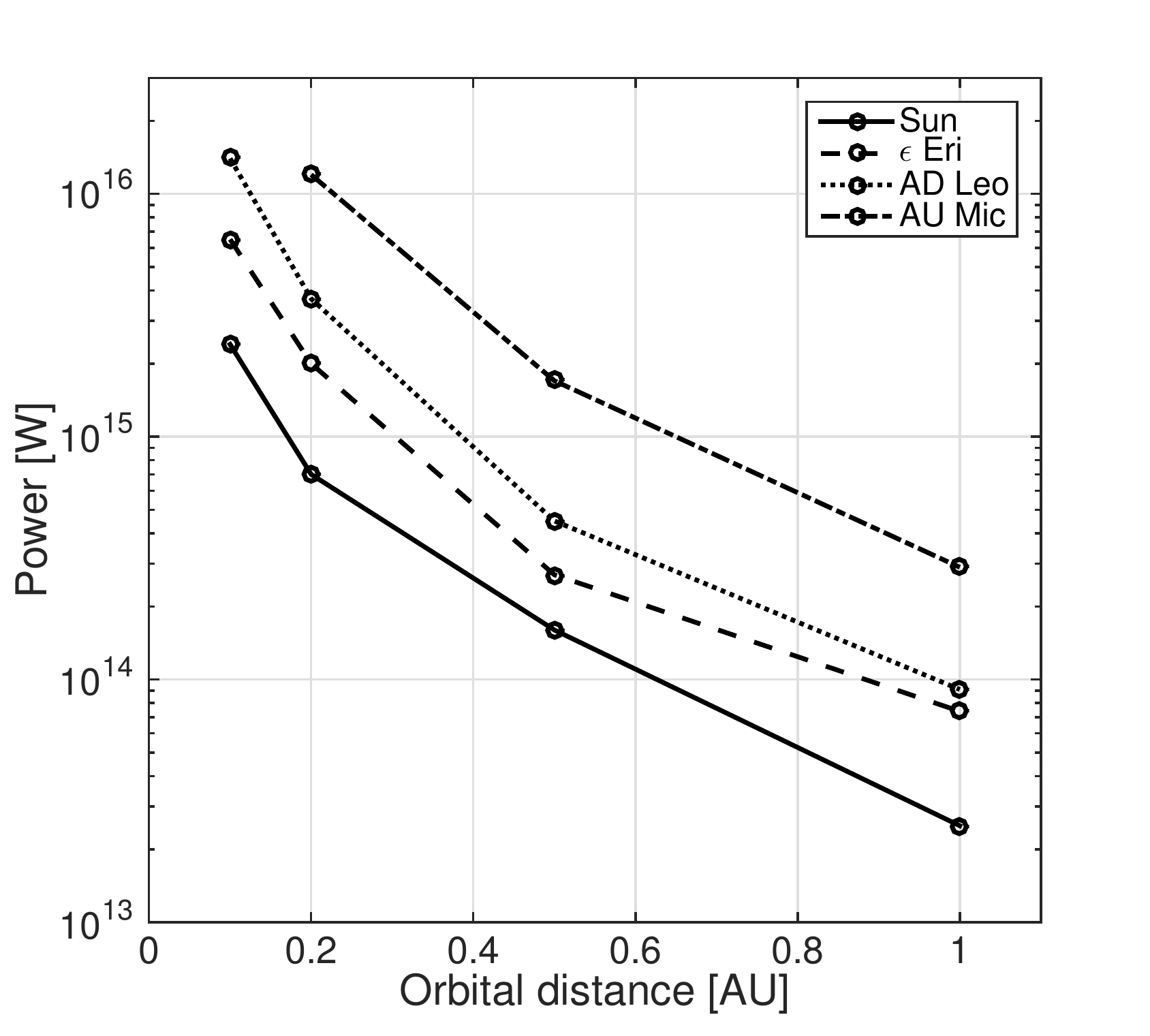}
\caption{Total H$_3^+$ emission power from EGPs orbiting various stars at different orbital distances.}
\label{fig:H3p_emission_power}
\end{figure}

We determined the total output power from the H$_3^+$ ion and the output in specific spectral lines for EGPs around a Sun-like star (Table~\ref{tab:H3p_power_sun}), $\epsilon$ Eri (Table~\ref{tab:H3p_power_eeri}), AD Leo (Table~\ref{tab:H3p_power_adleo}), and AU Mic (Table~\ref{tab:H3p_power_aumic}). The total output power is plotted in Fig.~\ref{fig:H3p_emission_power} for EGPs at different orbital distances around each star. The values given here were calculated using globally averaged H$_2$ and H$_3^+$ densities and considering emission from the planet over $2\pi$~steradians.

As a result of enhanced ionisation and hence increased H$_3^+$ column densities, the output power is higher around more active stars at a given orbital distance. For example, the total output from H$_3^+$ for a planet at 0.1~AU from AD Leo is $1.4\times10^{16}$~W, whereas for an EGP at the same distance from the Sun, it would be $2.4\times10^{15}$~W. Likewise, for a planet orbiting a given star, smaller orbital distances mean higher H$_3^+$ emissions. This is because our model predicts higher temperatures and H$_3^+$ column densities close to the star.

\citet{Koskinen2007a} determined H$_3^+$ emission powers for an EGP at different orbital distances around a solar-like star (HD209458) and obtained very similar values to our results for a planet orbiting the Sun. \citet{Yelle2004} calculated an output power of $9.9\times10^{16}$~W for HD209458b, an EGP orbiting at 0.047~AU, which is also consistent with our findings.

It would appear that the best targets for detecting H$_3^+$ emissions are planets orbiting close-in to active stars with high XUV irradiances (such as AD Leo or AU Mic). However, this may not be the case: in the closest planets orbiting the most active stars, we have found that the H$_3^+$ layer is pushed to higher pressures than for planets in the stable escape regime at larger orbital distances (see Sect.~\ref{sec:vary_a}). At these higher pressures, it is likely that the presence of heavy species (not yet included in our models) will destroy H$_3^+$. This means that in particular for EGPs at small orbital distances from AU Mic and AD Leo, the values of output power from H$_3^+$ presented here are likely overestimated. Therefore we did not determine H$_3^+$ emissions from an EGP at 0.1~AU from AU Mic.

Past observational campaigns have failed to detect these emissions from hot Jupiters. \citet{Shkolnik2006} observed six low-mass stars (of type F, G, K, and M) using the NASA Infrared Telescope Facility's Cryogenic Near-IR Facility Spectrograph (CSHELL). They determined emission limits for each observed system. The lowest limit they found is $6.3\times10^{17}$~W for GJ436b, a hot Jupiter orbiting an M dwarf located at 10.2~pc from Earth. According to our calculations, the emission limits set by the observations made by \citet{Shkolnik2006} are too high to detect H$_3^+$ emissions.

Measurements of planetary absorption line depths as low as $10^{-4}$ with respect to the stellar continuum are becoming possible with ground-based high-resolution spectrographs such as CRIRES on the VLT \citep[see e.g.][]{DeKok2013,Brogi2012}. Based on our model predictions for H$_3^+$ emissions, we estimate the planetary emission to stellar continuum contrast to be approximately $7\times 10^{-7}$ for a planet at 0.2~AU from a star similar to AD Leo, dropping to $3\times 10^{-8}$ at 1~AU from the same star. These estimates are based on the 3.534~$\mu$m R(3,3-) line, for a spectral resolution of about 100,000 and on the simplistic assumption that the planet's continuum emission can be approximated to first order by a black body at the equilibrium temperature. Therefore, even accounting for gains by observing multiple lines, the detection of H$_3^+$ emissions from EGPs is not possible with current technologies. The difficulties are further compounded by the relatively small number of H$_3^+$ emission lines (of the order of tens rather than thousands as in the case of CO absorption lines) that does not lead to the same gain in the cross-correlations. Furthermore, simulations by \citet{DeKok2013} suggested that signatures from emission lines might be more difficult to retrieve than from absorption lines (see their Fig.~5).

A number of recent studies (\citealt{Snellen2015,Kawahara2014}; see also \citealt{Sparks2002}) have suggested different methods to combine high spatial and high spectral resolution imaging techniques, with simulations suggesting contrast improvements of between a factor of 100 and 1000 depending on the technique and planet-star configuration. If these are indeed realised, line contrasts as low as $10^{-7}$ will become detectable and bring direct detections of H$_3^+$ in planetary systems such as the ones modelled here within our reach.

The presence of strong aurora may also increase the detectability of IR emissions from EGPs. There are currently no constraints on exoplanetary magnetospheres, but it is possible that these could induce emissions similar to the powerful auroral emissions seen at Jupiter \citep[e.g.][]{Radioti2013,Miller2006}.

\begin{table}
\centering
\caption{H$_3^+$ emission power for an EGP orbiting a Sun-like star. All units are Watts.}
\begin{tabular}{l|cccc}
\toprule
 & Sun & & & \\
 & 0.1~AU & 0.2~AU & 0.5~AU & 1~AU \\
\midrule
Q(1)	& $1.4\times10^{13}$	& $6.4\times10^{12}$	& $1.5\times10^{12}$	& $3.8\times10^{11}$ \\
Q(3)	& $9.5\times10^{12}$	& $4.9\times10^{12}$	& $1.2\times10^{12}$	& $3.0\times10^{11}$ \\
R(1)	& $1.1\times10^{13}$	& $5.2\times10^{12}$	& $1.3\times10^{12}$	& $3.2\times10^{11}$ \\
R(3)	& $1.5\times10^{13}$	& $6.9\times10^{12}$	& $1.7\times10^{12}$	& $4.1\times10^{11}$ \\
Total	& $2.4\times10^{15}$	& $7.0\times10^{14}$	& $1.6\times10^{14}$	& $2.5\times10^{13}$ \\
\bottomrule
\end{tabular}
\label{tab:H3p_power_sun}
\end{table}

\begin{table}
\centering
\caption{H$_3^+$ emission power for an EGP orbiting $\epsilon$ Eri. All units are Watts.}
\begin{tabular}{l|cccc}
\toprule
 & $\epsilon$ Eri & & & \\
 & 0.1~AU & 0.2~AU & 0.5~AU & 1~AU \\
\midrule
Q(1)	& $3.5\times10^{13}$	& $1.5\times10^{13}$	& $2.5\times10^{12}$	& $7.5\times10^{11}$ \\
Q(3)	& $2.4\times10^{13}$	& $1.1\times10^{13}$	& $2.0\times10^{12}$	& $5.6\times10^{11}$ \\
R(1)	& $2.7\times10^{13}$	& $1.2\times10^{13}$	& $2.1\times10^{12}$	& $6.0\times10^{11}$ \\
R(3)	& $3.9\times10^{13}$	& $1.7\times10^{13}$	& $2.7\times10^{12}$	& $8.1\times10^{11}$ \\
Total	& $6.5\times10^{15}$	& $2.0\times10^{15}$	& $2.7\times10^{14}$	& $7.4\times10^{13}$ \\
\bottomrule
\end{tabular}
\label{tab:H3p_power_eeri}
\end{table}

\begin{table}
\centering
\caption{H$_3^+$ emission power for an EGP orbiting AD Leo. All units are Watts.}
\begin{tabular}{l|cccc}
\toprule
 & AD Leo & & & \\
 & 0.1~AU & 0.2~AU & 0.5~AU & 1~AU \\
\midrule
Q(1)	& $7.0\times10^{13}$	& $2.8\times10^{13}$	& $4.7\times10^{12}$	& $9.9\times10^{11}$ \\
Q(3)	& $4.6\times10^{13}$	& $1.9\times10^{13}$	& $3.7\times10^{12}$	& $7.6\times10^{11}$ \\
R(1)	& $5.3\times10^{13}$	& $2.1\times10^{13}$	& $3.9\times10^{12}$	& $8.1\times10^{11}$ \\
R(3)	& $7.8\times10^{13}$	& $3.0\times10^{13}$	& $5.1\times10^{12}$	& $1.1\times10^{12}$ \\
Total	& $1.4\times10^{16}$	& $3.7\times10^{15}$	& $4.5\times10^{14}$	& $9.1\times10^{13}$ \\
\bottomrule
\end{tabular}
\label{tab:H3p_power_adleo}
\end{table}

\begin{table}
\centering
\caption{H$_3^+$ emission power for an EGP orbiting AU Mic. All units are Watts.}
\begin{tabular}{l|ccc}
\toprule
 & AU Mic & & \\
 & 0.2~AU & 0.5~AU & 1~AU \\
\midrule
Q(1)	& $6.7\times10^{13}$	& $1.4\times10^{13}$	& $2.7\times10^{12}$ \\
Q(3)	& $4.4\times10^{13}$	& $1.0\times10^{13}$	& $2.1\times10^{12}$ \\
R(1)	& $5.0\times10^{13}$	& $1.1\times10^{13}$	& $2.2\times10^{12}$ \\
R(3)	& $7.5\times10^{13}$	& $1.5\times10^{13}$	& $3.0\times10^{12}$ \\
Total	& $1.2\times10^{16}$	& $1.7\times10^{15}$	& $2.9\times10^{14}$ \\
\bottomrule
\end{tabular}
\label{tab:H3p_power_aumic}
\end{table}

\section{Conclusion}
This paper builds upon \citet{Chadney2015}, where the authors discussed the neutral atmosphere and escape regimes of EPGs around these stars. Here, we have studied the ionised region of EGP upper atmospheres by applying an ionospheric model to planets irradiated by XUV radiation from the Sun at solar minimum and maximum, as well as with the stars $\epsilon$ Eri, AD Leo and AU Mic. As in solar system gas giants \citep[e.g.][]{Moore2004,Moore2009}, we found that the dominant ions in the EUV-driven part of the ionosphere on EGPs are H$^+$ and H$_3^+$. In planets orbiting at large orbital distances, the upper atmosphere is in the `stable' regime (i.e. undergoing Jeans escape). In this situation, there is a region in the lower ionosphere where H$_3^+$ is the dominant ion species. Peak H$_3^+$ number densities are reached in layers where stellar soft X-ray fluxes are absorbed, whereas the peak in H$^+$ is located in the EUV heating layer. Thus the highest value of $n_{\text{H}_3^+}$ is determined by stellar soft X-ray flux levels and the highest value of $n_{\text{H}^+}$ is determined by stellar EUV flux levels. This is a noteworthy difference, since stellar fluxes in the X-ray and EUV bands scale differently with stellar activity \citep[see][]{Chadney2015}.

As the XUV stellar flux is increased, whether because of a more active star or a smaller orbital distance, atmospheric temperatures and flux levels become high enough for the planet to shift to a regime of hydrodynamic escape \citep[see][]{Chadney2015}. This leads to a change in temperature profile (a very large temperature peak is formed) and neutral density structure (molecular hydrogen is fully dissociated above a certain pressure level within the ionosphere). In this situation, the H$_3^+$ peak is pushed to higher pressures, mainly as a result of the increased photo- and electron-impact ionisation of H$^+$ at higher pressure levels (H$_3^+$ is destroyed though recombination with the increased number of electrons). At high enough stellar flux levels, H$^+$ is the dominant ion throughout the modelled region of atmosphere. Any peak in H$_3^+$ would then be confined to pressures below the lower boundary of the model, where it is likely destroyed though reactions with heavy species (e.g. water, hydrocarbons). 

We here calculated the photo-electron energy degradation to determine ionisation by photo-electrons. This has not been included before in EGP studies, and we showed that this considerably affects ionisation below 10~nbar, where it is the dominant form of ionisation, pushing the ionosphere to lower altitudes than previously assumed.

The H$_3^+$ IR emissions we predict from EGPs around active stars are higher than for planets orbiting at the same distance from a star of similar age to the Sun. However, the emissions may still be too low to be detected by the current technologies. The past focus on detecting these emissions from hot Jupiters might not be the optimum strategy. It relies on the assumption that closer orbiting EGPs will have higher column densities of H$_3^+$ and higher temperatures, leading to higher emissions. This assumption is valid in a pure H/H$_2$/He atmosphere, such as that considered in this study. However, given that for EGPs orbiting very close to their stars, we predict that H$_3^+$ is confined to a layer at the bottom of the ionosphere, possibly below the homopause, it is likely destroyed by heavy species. The location of the homopause is dependent on the value of the eddy diffusion coefficient, which is not well known in exoplanets. The predictions of H$_3^+$ densities in our simulations rely on the homopause being located at a pressure of 1~$\mu$bar, meaning an eddy diffusion coefficient $K_{zz}$ lower than $\sim10^2$ -- $10^3$~m$^2$~s$^{-1}$. Higher values of $K_{zz}$ will mean that hydrocarbons could be present in the H$_3^+$ layer. Therefore a detection of H$_3^+$ emissions would also place constraints on $K_{zz}$.

Given the low levels of H$_3^+$ emission that we predict, these may be difficult to detect with current telescopes, even with planets around active stars. However, new techniques involving high spatial and high spectral resolution may provide sufficient planet-to-star contrast to allow a detection of H$_3^+$ emissions. Despite this, other diagnostics of EGP ionospheres could prove to be more promising in the near future, such as the detection of radio emissions from magnetosphere-ionosphere coupling \citep{Nichols2011}. Although these observations have yet to succeed for EGPs, radio emissions emanating from a brown dwarf have recently been detected \citep{Hallinan2015}.

\begin{acknowledgements}
J.M.C., M.G., and Y.C.U.\ are partially funded by the UK Science \& Technology Facilities Council (STFC) through the Consolidated Grant to Imperial College London. J.M.C.\ has also received support from SFTC through a postgraduate studentship, award number ST/J500616/1. T.T.K. acknowledges support from the National Science Foundation (NSF) grant AST 1211514. J.S.F. acknowledges support from the Spanish MINECO through grant AYA2011-30147-C03-03.
\end{acknowledgements}

\bibliographystyle{aa}
\bibliography{references}

\end{document}